%% file: main.tex
\documentclass[journal,twoside,web]{ieeecolor}

\usepackage{generic}
\usepackage{cite}
\usepackage{amsmath,amssymb,amsfonts}
\usepackage{algorithmic}
\usepackage{graphicx}
\usepackage{algorithm,algorithmic}
\usepackage{hyperref}
\hypersetup{hidelinks=true}
\usepackage{textcomp}

\usepackage{empheq}

\def\BibTeX{{\rm B\kern-.05em{\sc i\kern-.025em b}\kern-.08em
    T\kern-.1667em\lower.7ex\hbox{E}\kern-.125emX}}
\let\labelindent\relax 

\usepackage{booktabs}
\usepackage{cuted}
\usepackage{enumitem}
\usepackage{mathtools}
\usepackage{multirow}
\usepackage{physics}
\usepackage{subcaption}

\usepackage{makecell}
\usepackage{threeparttable}
\usepackage{pifont}

\newcommand{\xmark}{\ding{55}}

\newcommand{\R}{\mathbb{R}}

\newtheorem{definition}{Definition}
\newtheorem{remark}{Remark}
\newtheorem{example}{Example}

\newtheorem{theorem}{Theorem}
\newtheorem{proposition}{Proposition}
\newtheorem{lemma}{Lemma}
\newtheorem{corollary}{Corollary}

\makeatletter
\newcommand{\customlabel}[2]{%
  \def\@currentlabel{#1}%
  \label{#2}%
}
\makeatother

\begin{document}
\bstctlcite{IEEEexample:BSTcontrol}
%
%
\title{React to Surprises: Stable-by-Design Neural Feedback Control and the Youla-REN}

\author{Nicholas H. Barbara, Ruigang Wang, Alexandre Megretski, and Ian R. Manchester
\thanks{This work was supported in part by the Australian Research Council (DP230101014 and IH210100030) and Google LLC.}\thanks{NB, RW, and IM are with the Australian Centre for Robotics (ACFR) and the School of Aerospace, Mechanical and Mechatronic Engineering, The University of Sydney, Australia,
{\tt\small \{nicholas.barbara, ruigang.wang, ian.manchester\}@sydney.edu.au}. AM is with the Laboratory for Information and Decision Systems, Dept. Electrical Engineering and Computer Science, Massachusetts Institute of Technology, {\tt\small ameg@mit.edu}.}
}
\maketitle

%
%
\begin{abstract}
We study parameterizations of stabilizing nonlinear policies for learning-based control. We propose a structure based on a nonlinear version of the Youla-Ku\v{c}era parameterization combined with robust neural networks such as the recurrent equilibrium network (REN). The resulting parameterizations are unconstrained, and hence can be searched over with first-order optimization methods, while always ensuring closed-loop stability by construction. We study the combination of (a) nonlinear dynamics, (b) partial observation, and (c) incremental closed-loop stability requirements (contraction and Lipschitzness). We find that for the combination of (c) with either (a) or (b), a contracting and Lipschitz Youla parameter always leads to contracting and Lipschitz closed loops. However, if all three hold, then incremental stability can be lost with exogenous disturbances. Instead, a weaker condition is maintained, which we call d-tube contraction and Lipschitzness. We further obtain converse results showing that the proposed parameterization covers all contracting and Lipschitz closed loops for certain classes of nonlinear systems. Numerical experiments illustrate the utility of our parameterization when learning controllers with built-in stability certificates for: (i) ``economic'' rewards without stabilizing effects; (ii) short training horizons; and (iii) uncertain systems.
\end{abstract}

\begin{IEEEkeywords}
    Youla parameterization, recurrent equilibrium networks, contraction, IQC, reinforcement learning
\end{IEEEkeywords}

%
%

%
%
\section{Introduction} \label{sec:intro}

A conceptually simple yet powerful approach to control design is to first (a) parameterize a family of candidate feedback policies, and then (b) search over policy parameters to (locally) optimize some cost or reward function. This general approach is central to deep reinforcement learning (RL), which has successfully been applied in many complex domains from games \cite{Silver++2017} to robotics \cite{tang2025deep} to nuclear fusion \cite{Degrave++2022}, among others.

In deep RL it is Step (b), the search for policy parameters, which has been the principal focus \cite{Sutton+Barto2018}. The main challenge is estimating the gradient of the cost with respect to parameters when only noisy observations of states and rewards are available. Policy gradient methods randomly perturb actions or parameters and estimate gradients via Monte Carlo methods, likelihood ratios \cite{williams1992simple}, or stochastic approximation methods \cite{meyn2022control}. In actor-critic methods such as proximal policy optimization (PPO) \cite{Schulman++2017}, a value function is also estimated to reduce variance of gradient estimates. More recently, RL methods have taken advantage of differentiable physics simulators through which policy gradients can be computed algorithmically \cite{xu2022accelerated, suh2022differentiable, Freeman++2021, howell2022dojo}.

In this paper we study Step (a), or choosing a policy parameterization, which has seen relatively little focus in RL. Since the standard RL problem statement includes no prior knowledge of the dynamics, black-box policy families such as deep neural networks are commonly used with the particular architecture and hyperparameters chosen based on experience and empirical tuning. Such policy parameterizations are highly expressive, due to the universal approximation properties of neural networks, but it is usually difficult or impossible to \textit{certify} important properties of the closed-loop system such as stability, robustness, or safety, especially during training \cite{Manchester++2026}.

In the control literature there is a long history of research into parameterizations of stabilizing controllers (see Sections \ref{sec:intro_youla}, \ref{sec:youla-review}), which this paper builds on. However, we argue below that most classical results do not meet the needs of modern learning-based control paradigms such as deep RL, in that the problem settings are too restrictive and the proposed solutions are not directly compatible with the standard machine learning tool-chain based on automatic differentiation.

\subsection{Objectives of this Paper}\label{sec:intro-objectives}

The main objective of this paper is to help bridge the gap between control theory and deep RL by introducing policy parameterizations that satisfy the following requirements.

\begin{enumerate}[label=\textbf{R\arabic*}]
    \item \textbf{Stable by design:} All controllers in the policy class guarantee robust closed-loop stability, of a form suitable for learning-based control of nonlinear, partially-observed systems, over a wide range of operating conditions. \label{req:stable}
    \item \textbf{Expressive:} Subject to requirement \ref{req:stable}, the policy class is as flexible as possible to maximize coverage of high-performance policies, i.e., to reduce as much as possible the conservatism of the stability constraints.\label{req:expressive}
    \item \textbf{Differentiable:} The control policy is smoothly parameterized via a vector in $\mathbb R^N$, without requiring auxiliary constraints for stability, thus enabling the use of standard policy gradient methods. \label{req:differentiable}
 \end{enumerate}
By embedding closed-loop guarantees within the policy class, we decouple stability from the choice of optimization algorithm, reward functions, and data used to train the policy. In a sense, our overall goal is to maximize policy expressivity subject to stability and differentiability.

\subsection{Black Box vs Structured Neural Feedback Policies}\label{sec:intro_youla}

\begin{figure}[!t]
    \centering
    \begin{subfigure}[b]{0.272\linewidth}
        \centering
        \includegraphics[trim={6cm 13.4cm 18.2cm 1.4cm},clip,width=\linewidth]{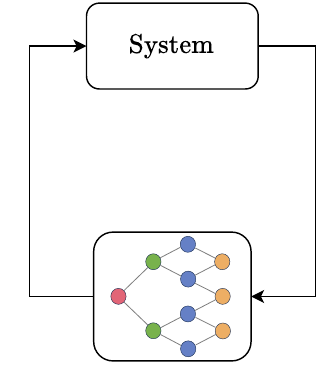}
        \caption{Black-box.}
        \label{fig:feedback-architecture}
    \end{subfigure}
    \hfill
    \begin{subfigure}[b]{0.292\linewidth}
        \centering
        \includegraphics[trim={11.8cm 13.4cm 12cm 1.4cm},clip,width=\linewidth]{Images/youla_high_level_v6.drawio.pdf}
        \caption{Residual.}
        \label{fig:residual-architecture}
    \end{subfigure}
    \hfill
    \begin{subfigure}[b]{0.405\linewidth}
        \centering
        \includegraphics[trim={17.8cm 13.4cm 3.7cm 1.4cm},clip,width=\linewidth]{Images/youla_high_level_v6.drawio.pdf}
        \caption{Youla-REN.}
        \label{fig:youla-architecture}
    \end{subfigure}
    \caption{Neural feedback control architectures.}
    \label{fig:rl-architectures}
\end{figure}

In this paper we propose structured policy architectures that use prior knowledge to embed stability guarantees in the controller parameterization. Fig.~\ref{fig:feedback-architecture} shows the standard RL paradigm of a learned black-box policy (e.g., neural network) in feedback with a system (\textit{a.k.a.} environment) \cite{Sutton+Barto2018}. The black-box architecture does not assume any prior knowledge, but subsequently cannot guarantee any closed-loop stability or robustness properties.

Fig.~\ref{fig:residual-architecture} shows the residual RL architecture \cite{Silver++2019,Johannik++2019,Luo++2025}, which utilizes prior knowledge in the form of a known (e.g., model-based) stabilizing ``base controller'' by augmenting it with a learned neural network. The purpose of the base controller is to bootstrap performance, and while it may be possible to certify stability with a neural network via the small gain theorem, this will generally be conservative and closed-loop stability and robustness are usually not enforced in residual RL.

Fig.~\ref{fig:youla-architecture} shows the architecture we use in this paper, which can reduce or even eliminate this conservatism if further prior knowledge in the form an observer (\textit{a.k.a.} state estimator) is available. Observers can be thought of as ``model-pass filters,'' and the difference between their predicted measurement and the actual measurement represents ``surprises'' (\textit{a.k.a} innovations). Inspired by the Youla parameterization \cite{Youla++1976, Anderson1998, Mahtout2020}, this architecture ``reacts to surprises'' through a so-called \textit{Youla parameter}, which can be any stable system and is treated as a free parameter. Its main attraction is that under certain conditions reviewed below: (1) a stable Youla parameter implies closed-loop stability; and (2) it is expressive in that it covers \textit{all} stabilizing controllers.

\subsection{Parameterizations of Stabilizing Control Policies}

Requirements \ref{req:stable} and \ref{req:expressive} in Section~\ref{sec:intro-objectives} call for a rich (ideally complete) parameterization of stabilizing controllers to ensure stability without unduly compromising performance. 

For linear systems, essentially all notions of stability coincide and the parameterization of stabilizing controllers has a long history. Concurrent work of Youla \cite{Youla++1976} and \cite{Kucera1975} was the first to parameterize all stabilizing linear controllers for general (possibly unstable) multi-variable linear systems -- see also the surveys \cite{Anderson1998, Mahtout2020}. Roughly speaking, given one stabilizing controller, the Youla parameterization expresses all others in terms of a stable ``Youla parameter'' $Q$, which itself is a free stable linear system. It has long played a central role in robust  and adaptive control theory \cite{Zhou++1996,Anderson1998}, and was first proposed in the (linear) RL context in \cite{Roberts++2011}. It was initially introduced in terms of co-prime transfer matrix factorizations, but can also be expressed in a state-space innovations-feedback form in which $Q$ ``reacts to surprises'' as in Fig.~\ref{fig:youla-architecture} \cite{Doyle1984}. Similar parameterizations have been proposed under many names over the years, including: $\mathcal{Q}$-parameterization \cite{Zames1981}; internal model control (IMC) \cite{Garcia+Morari1982}; disturbance feedback in model predictive control (MPC) \cite{Goulart++2006}; and ``nature's y'' in online learning \cite{simchowitz2020improper}. 

Extensions of the Youla parameterization to nonlinear systems have been studied for many years
\cite{Desoer+Liu1982,Desoer+Lin1983, sontag1989smooth, Tay+Moore1998,Paice+Moore1990,Paice+vanderSchaft1996,Fujimoto+Sugie2000,Lu1995,Imura+Yoshikawa1997}. The basic theory in terms of co-prime factors was established in \cite{sontag1989smooth, Tay+Moore1998,Paice+Moore1990}, however constructive tools were lacking. Kernel representations \cite{Paice+vanderSchaft1996, Fujimoto+Sugie2000} made connections with nonlinear state-variable controller and observer design, leading to more constructive tools e.g. \cite{Lu1995,Imura+Yoshikawa1997}.

A limitation of \cite{Desoer+Liu1982,Desoer+Lin1983, sontag1989smooth,Paice+Moore1990,Paice+vanderSchaft1996,Fujimoto+Sugie2000,Lu1995,Imura+Yoshikawa1997} is the focus on asymptotical or finite-gain stabilization about a known equilibrium. We argue that in RL and learning-based control, this is insufficient since the objective is usually to obtain performant controllers over a wide range of operating conditions, possibly unseen in the training data. We argue that \emph{incremental} stability notions such as contraction \cite{Lohmiller+Slotine1998} are more appropriate, since they ensure stable behavior when generalizing to unseen data.

\subsection{Direct Parameterizations for Differentiable Policies}

Requirement \ref{req:differentiable} in Sec. \ref{sec:intro-objectives} ensures compatibility with the standard machine learning tool-chain based on automatic differentiation and gradient methods for optimization.
Here we argue that this is an additional benefit of the Youla parameterization, in that it parameterizes the control policy as a smooth transformation of a stable, but otherwise free, dynamical system $Q$. If a rich set of stable dynamical systems can in turn be smoothly parameterized over $\mathbb R^N$ for some $N$, then we have a so-called ``direct'' parameterization \cite{Wang+Manchester2023} of stabilizing controllers.

Prior work introduced convex parameterizations of contracting and Lipschitz nonlinear dynamic models based on polynomials \cite{tobenkin2010convex, tobenkin2017convex} and neural networks \cite{revay2020convex}. However, stability constraints were expressed in the form of linear matrix inequalities which limited their scalability and required custom optimization methods \cite{umenberger2018specialized,revay2020convex}.

A key recent advance was the direct (i.e., unconstrained, smooth, surjective) parameterization of rich classes of Lipschitz neural networks \cite{Wang+Manchester2023} and contracting and Lipschitz dynamic systems called Recurrent Equilibrium Networks (RENs) \cite{Revay++2023}.
By composing the direct parameterization of RENs with the Youla parameterization, we can construct a direct parameterization of stabilizing controllers, illustrated below:
\begin{equation}
\mathbb R^N \,\xrightarrow{\textrm{REN param}}\, \textrm{Stable }\mathcal Q \,\xrightarrow{\textrm{Youla param}}\, \textrm{Stabilizing }\mathcal K.
\end{equation}

\subsection{Contributions \& Paper Outline}

The main contributions of this paper are:
\begin{itemize}
    \item We introduce a version of the Youla parameterization in the general case of partially-observed nonlinear systems with disturbances, and show that loss of contraction and Lipschitzness is possible in this case.

    \item We introduce the notion of \textit{d-tube contraction and Lipschitzness} -- i.e., contracting and Lipschitz responses to known inputs and bounded response to unknown inputs -- and show that our parameterization satisfies this property.
    
    \item We study the coupling between innovations and control augmentations, and show how closed-loop contracting and Lipschitz responses to disturbances can be achieved with Lipschitz-bounded Youla parameters.

    \item We provide partial converse results proving that the parameterization contains \emph{all} contracting and Lipschitz closed-loops for certain classes of nonlinear systems.

    \item Combined with RENs, we show via numerical experiments that our resulting \emph{Youla-REN} policies can achieve strong performance on deep RL tasks while preserving closed-loop stability under disturbances and uncertainty.
\end{itemize}
As such, this paper significantly extends earlier work by the authors for linear systems \cite{Wang+Manchester2022, Wang++2022, Revay++2023} and partially-observed nonlinear systems without disturbances \cite{Barbara++2023a}.

Concurrent work in \cite{Kawano++2024} also extends the Youla parameterization to the contraction framework, and includes similar statements to  Theorem~\ref{thm:partial-converse} in the present paper, variants of which also appeared in \cite{Barbara++2023a}. However, \cite{Kawano++2024} was limited to the disturbance-free setting.
Further concurrent work \cite{Furieri++2022,Furieri++2024,Galimberti++2024} studies the Youla parameterization from the perspective of input-output operators, and also utilizes RENs as the stable parameter. However, \cite{Furieri++2022,Furieri++2024} only consider state-feedback settings, and \cite{Galimberti++2024} is limited to finite-gain stability for stable plants and assumes disturbances are measurable for the results on closed-loop contraction and Lipschitzness. We summarise the connection between these works and our paper in Table~\ref{tab:youla-theory}.

The remainder of this paper is organized as follows. Section~\ref{sec:problem-formulation} formulates the problem of constructing robustly-stabilizing policy parameterizations. Section~\ref{sec:youla-review} introduces a nonlinear version of the Youla parameterization. Our main theoretical results are presented in Sections~\ref{sec:minitheory}--\ref{sec:converse}. Section~\ref{sec:exp} studies the Youla-REN via numerical examples. Conclusions and proofs are provided in Section~\ref{sec:conc} and Appendices~\ref{app:proof-main-result}--\ref{app:proof-partial-converse}.

\begin{table}[!t]
    \caption{Comparison of studies on nonlinear Youla parameterizations across problem settings. We examine the general case of nonlinear (NL), partially-observed (PO) systems with incremental stability (contracting and Lipschitz closed loops). Non-incremental stability refers to finite-gain stability. See Section~\ref{sec:youla-review} and Fig.~\ref{fig:youla-sys} for definitions of $\mathcal{G}, \mathcal{K}, \mathcal{K}_\mathcal{Q}$.}
    \label{tab:youla-theory}
    \centering
    \begin{threeparttable}
    \setlength{\tabcolsep}{4pt}
    \begin{tabular}{cccccc}
    \toprule
    \thead{\textbf{NL} $\mathcal{G}$} & 
    \thead{\textbf{PO} $\mathcal{G}$} & 
    \thead{\textbf{Incremental} \\ \textbf{stability}} & 
    \thead{$\mathcal{K} = \mathcal{K}_\mathcal{Q} \iff$\\ $\mathcal{K}$ \textbf{stabilizes} $\mathcal{G}$} & \thead{\textbf{Section}} &
    \thead{\textbf{References}} \\
    \midrule
    \xmark      & \checkmark & \checkmark & \checkmark & \ref{sec:theory-linear} & \cite{Wang++2022} \\
    \checkmark  & \xmark     & \checkmark & \checkmark & \ref{sec:theory-state-feedback} & This paper\\
    \checkmark  & \checkmark & \xmark     & \checkmark & \ref{sec:theory-finite-gain} & \cite{Fujimoto+Sugie2000,Imura+Yoshikawa1997} \\
    \midrule
    (\checkmark)\tnote{*}  & \xmark                     & \xmark      & \checkmark        & -- & \cite{Furieri++2024} \\
    \checkmark  & \xmark     & \xmark     & \checkmark & -- & \cite{Furieri++2022} \\
    (\checkmark)\tnote{*}  & (\checkmark)\tnote{$\dagger$} & \checkmark  & \checkmark        & -- & \cite{Galimberti++2024} \\
    \checkmark          & (\checkmark)\tnote{$\ddagger$}  & \checkmark      & (\checkmark)\tnote{$^\triangledown$} & -- & \cite{Kawano++2024} \\ 
    \midrule
    \checkmark  & \checkmark & \checkmark & (\checkmark)\tnote{\S} & \ref{sec:theory}, \ref{sec:converse} & This paper \\
    \bottomrule
    \end{tabular}
    \begin{tablenotes}
        \item[*] Assumes $\mathcal{G}$ is stable.
        \item[$\dagger$] Assumes disturbances are known and exactly measurable.
        \item[$\ddagger$] Assumes zero disturbances.
        \item[$\triangledown$] Contraction with transients due to initial conditions (see Sec.~\ref{sec:d-tube}); converse results only for closed loops with constant contraction metrics.
        \item [\S] Under the conditions in Theorems~\ref{thm:youla-disturbance-tube}--\ref{thm:partial-converse}.
    \end{tablenotes}
    \end{threeparttable}
\end{table}

\subsection{Notation} \label{sec:intro-notation}

Let $\mathbb{N}$ and $\bigl(\R^+\bigr)$ $\R$ be the set of natural and (positive) real numbers, respectively. We denote the set of sequences $x:\mathbb{N}\rightarrow \mathbb{R}^n$ by $\ell^n$, where the superscript $n$ is omitted if it is clear from context. Let $\ell_2^n\subset \ell^n$ be the set of sequences with finite $\ell_2$ norm, i.e., $\|x\|:=\left(\sum_{t=0}^{\infty}|x_t|^2\right)^{\frac{1}{2}} < \infty$ for any $x\in \ell_2^n$ where $|\cdot|$ is the Euclidean norm. We use $\|x\|_T:=\bigl(\sum_{t=0}^{T}|x_t|^2\bigr)^{\frac{1}{2}}$ to denote the truncated norm with $T\in\mathbb{N}$, and $\|x\|_{T,\infty}:=\max_{0\le s\le T} |x_s|$ for the truncated infinity norm. We write $A(\succeq) \succ 0$ for positive (semi-)definite matrices. A function $f:\mathbb{R}^m\rightarrow\mathbb{R}^n$ is said to be $\gamma$-Lipschitz with $\gamma\in \R^+$ if $|f(a)-f(b)|\leq \gamma|a-b|$ for all $a,b\in \mathbb{R}^n$. We denote the concatenation of vectors $u\in\mathbb{R}^{n}$, $v\in\mathbb{R}^m$ as $[u;v]$.

%
%
\section{Problem Formulation} \label{sec:problem-formulation}

\subsection{Problem Statement} \label{sec:prob-statement}
We consider nonlinear systems $\mathcal{G}$ of the form
\begin{equation} \label{eqn:system}
\mathcal{G}: \;\left\{
    \begin{aligned}
        x_{t+1}&=f(x_t, \eta_t, u_t) + w_t \\
        y_t &= h(x_t) + v_t
    \end{aligned}\right.
\end{equation}
with states $x_t \in \mathbb{R}^{n_x}$, inputs $u_t \in \mathbb{R}^{n_u}$, and outputs $y_t \in \mathbb{R}^{n_y}$. The states and measurements are perturbed by (unknown) additive process disturbances $w_t\in \mathbb{R}^{n_x}$ and measurement noise $v_t\in \mathbb{R}^{n_y}$, respectively. Here $\eta_t \in \R^{n_{\eta}}$ are known exogenous inputs, e.g. reference signals, feedforward commands, or disturbance previews. We denote 
the unknown disturbances $d_t = [w_t; v_t]$ and the controlled outputs $z_t = [x_t; u_t]$.

Our task is to parameterize stabilizing feedback policies $u = \mathcal{K}(\eta, y)$ for \eqref{eqn:system}. Specifically, our goal is to construct a \textit{policy class} parameterizing the set of controllers $\mathcal{K}_\theta$, where $\theta \in \mathbb{R}^N$ is a learnable parameter, which satisfies the three requirements in Section \ref{sec:intro-objectives}. Requirements \ref{req:stable} and \ref{req:differentiable} are hard requirements for learning-based control with robust stability guarantees. Requirement \ref{req:stable} ensures the policies are \textit{always} closed-loop stabilizing independently of the data or optimization algorithms. Requirement \ref{req:differentiable} ensures that the policy class is compatible with standard tools in machine learning based on unconstrained optimization. Requirement \ref{req:expressive} is a soft requirement  that the stability constraints are not overly restrictive, so it is possible to learn high-performing controllers for a wide range settings. 

With robust stability inherent to the policy class, we can learn stabilizing controllers $\mathcal{K}_\theta$ that seek to minimize arbitrary cost functions. A typical cost structure in deep RL is
\begin{equation} \label{eqn:general-cost-func}
    J = \mathop{\mathbb{E}}_{x_0,\eta,d} \left[ \sum_{t=0}^{T-1} g(x_t, u_t) + g_f(x_T) \right]
\end{equation}
for some $T\in\mathbb{N}$, where $g,g_f$ are any piecewise differentiable stage and terminal costs, respectively.

\subsection{Definitions of Stability \& Robustness} \label{sec:prob-definitions}

Consider a nonlinear system $u \mapsto y$ of the form
\begin{equation}\label{eqn:generic-system} 
    x_{t+1}=F(x_t,u_t),\quad y_t=H(x_t,u_t),
\end{equation}
with states $x_t\in \R^{n}$, inputs $u_t\in \R^{m}$, and outputs $y_t\in \R^p$. $F,H$ are assumed to be locally Lipschitz.

\begin{definition}[Contraction \cite{Lohmiller+Slotine1998}]\label{dfn:contraction}
    The system \eqref{eqn:generic-system} is said to be \emph{contracting} if for any two initial states $x_0^a, x_0^b \in \mathbb{R}^n$, given the same input sequence $u\in \ell^m$, the corresponding state trajectories $x^a,x^b\in \ell^n$ satisfy
    \begin{equation} \label{eqn:incremental-exp-stab}
        |x^a_t - x^b_t| \le \beta \alpha^t \abs{x_0^a - x_0^b}\quad \forall \, t \in \mathbb{N}
    \end{equation}
    with overshoot $\beta\in \R^+$ and contraction rate $\alpha \in [0,1)$.
\end{definition}

\begin{definition}[Incremental IQC \cite{Megretski+Rantzer1997}]\label{dfn:IQC}
    The system \eqref{eqn:generic-system} is said to satisfy the incremental \emph{integral quadratic constraint} (IQC) defined by a quadratic function $\sigma(\Delta u, \Delta y)$ which is negative definite \textit{w.r.t.} $\Delta y$ if any two trajectories $(x^a,u^a,y^a),(x^b,u^b,y^b)$ satisfy
    \begin{equation} \label{eqn:iqc}
        \sum_{t=0}^{T} 
        \sigma(\Delta u_t,\Delta y_t)
        \ge -\kappa(x^a_0, x^b_0)
        \quad  \forall T\in\mathbb{N}
    \end{equation}
    for some function $\kappa(x^a_0, x^b_0) \ge 0$ with $\kappa(x_0, x_0) = 0 \ \forall\,x_0\in\R^n$, and where $\Delta y_t = y^b_t - y^a_t$, $\Delta u_t = u^b_t - u^a_t$.
\end{definition}

An important special case of \eqref{eqn:iqc} is $\sigma(\Delta u,\Delta y)=\gamma^2|\Delta u|^2-|\Delta y|^2$ for some $\gamma\in\mathbb{R}^+$, which defines a Lipschitz bound, \textit{a.k.a.} incremental $\ell_2$-gain bound. 
Systems with small Lipschitz bounds are smooth in the sense that small changes to their inputs will not induce large variations in their outputs.

\begin{definition}[Lipschitz system]\label{dfn:lipschitz}
    The system \eqref{eqn:generic-system} is said to be \emph{Lipschitz} if there exists a $\gamma\in\R^+$ such that
    \begin{equation} \label{eqn:lipschitz}
        \norm{\Delta y}_{T} \le \gamma \norm{\Delta u}_{T} + \kappa(x^a_0, x^b_0) \quad \forall T\in\mathbb{N}
    \end{equation}
    with $\kappa(\cdot, \cdot), \Delta y, \Delta u$ as defined in Definition~\ref{dfn:IQC}. 
\end{definition} 

Many existing approaches to parameterizing stabilizing controllers are based on weaker notions of stability, such as finite $\ell_2$-gain stability. In the following definition, we assume $F(0,0)=0$ and $H(0,0)=0$ without loss of generality.

\begin{definition}[Finite-gain stability] \label{dfn:l2-stability}
    The system \eqref{eqn:generic-system} is said to be \textit{finite-gain stable} if there exists a $\gamma\in\mathbb{R}^+$ such that for any initial state $x_0\in\R^{n}$, inputs $u \in \ell^{m}$, and corresponding outputs $y \in \ell^p$, we have
    \begin{equation}
        \norm{y}_{T} \le \gamma \norm{u}_{T} + \kappa(x_0) \quad \forall T\in\mathbb{N}
    \end{equation}
    where $\kappa(x_0) \ge 0$ and $\kappa(0) = 0$.
\end{definition}

%
%
\section{Nonlinear Youla Parameterization} \label{sec:youla-review}

\begin{figure}[t]
    \centering
    \includegraphics[width=0.65\linewidth]{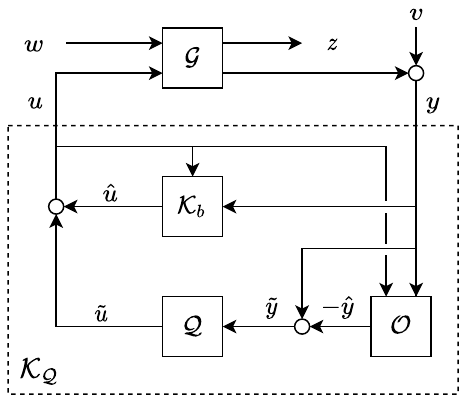}
    \caption{A version of the Youla-Ku\v{c}era parameterization. The Youla parameter $\mathcal{Q}$ augments a feedback controller $\mathcal{K}_b$ to control a plant $\mathcal{G}$. It reacts to difference $\tilde{y}$ between the measurements $y$ and predictions $\hat{y}$ from an observer $\mathcal{O}$. Known exogenous inputs $\eta$ have been omitted to simplify the diagram.}
    \label{fig:youla-sys}
\end{figure}

In this section, we construct the specific version of the Youla parameterization that we use. We assume that a stabilizing ``base'' controller has already been designed for the plant -- or as a special case, where the plant itself is stable -- and we wish to improve its closed-loop performance without compromising stability. Referring to Fig.~\ref{fig:youla-sys}, we learn a dynamical system $\mathcal{Q}$ called the \textit{Youla parameter} which adds controls $\tilde{u}$ to the output of the base controller $\mathcal{K}_b$. The total control signal is then $u = \hat{u} + \tilde{u}$. We assume that $\mathcal{K}_b$ takes the form
\begin{equation} \label{eqn:base-controller}
\mathcal{K}_b: \; \left\{
    \begin{aligned}
        s_{t+1} &= f_b(s_t, \eta_t, u_t, y_t) \\
        \hat{u}_t &= k(s_t, \eta_t, y_t)
    \end{aligned}\right.
\end{equation}
where $s_t \in \R^{n_s}$ is some internal state.
The first key ingredient in our parameterization is an observer $\hat{y} = \mathcal{O}(\eta, u, y)$ of the form
\begin{equation} \label{eqn:observer}
\mathcal{O}:\; \left\{
\begin{aligned}
\hat{x}_{t+1} &= f_o(\hat{x}_t, \eta_t, u_t, y_t) \\
\hat y_t &= h(\hat x_t)
\end{aligned}\right.
\end{equation}
where $\hat{x}_t \in \mathbb{R}^{n_x}$ is the state estimate and $\hat y_t\in\R^{n_y}$ is the output estimate. The \textit{innovations} sequence is
\begin{equation} \label{eqn:innovations}
    \tilde{y}_t := y_t - \hat y_t.
\end{equation}
Intuitively, the role of the observer is to decompose the measurements $y=\hat y+\tilde y$ into predictable measurements $\hat{y}$ and ``surprises'' $\tilde{y}$.
The second key ingredient is the Youla parameter $\tilde{u} = \mathcal{Q}(\eta, \tilde{y})$, which acts on the innovations, ``reacting to surprises''  as well as known exogenous inputs $\eta$. $\mathcal{Q}$ is itself a nonlinear dynamical system of the form
\begin{equation} \label{eqn:youla-param}
 \mathcal{Q}: \;   \left\{
 \begin{aligned}
     q_{t+1} &= f_q(q_t, \eta_t, \tilde{y}_t)\\
    \tilde{u}_t &= h_q(q_t, \eta_t, \tilde{y}_t)
 \end{aligned}\right.
\end{equation}
with internal state $q_t\in\mathbb{R}^{n_q}$ and augmenting control $\tilde{u}_t\in \R^{n_u}$. The resulting augmented controller $u=\mathcal{K}_{\mathcal{Q}}(\eta, y)$ is 
\begin{equation} \label{eqn:youla-ctrl}
\mathcal{K}_{\mathcal{Q}}:\;\left\{
    \begin{aligned}
        s_{t+1} &= f_b(s_t, \eta_t, u_t, y_t) \\
        \hat{x}_{t+1} &= f_o(\hat{x}_t, \eta_t, u_t, y_t) \\
        q_{t+1} &= f_q(q_t, \eta_t, y_t-h(\hat x_t)) \\
        u_t &= k(s_t, \eta_t, y_t) + h_q(q_t, \eta_t, y_t-h(\hat x_t)).
    \end{aligned}\right.
\end{equation}

\begin{remark}
    If external inputs $\eta$ are not provided to the Youla parameter such that $\tilde{u} = \mathcal{Q}(\tilde{y})$, then the Youla parameter \textit{only} reacts to surprises -- i.e., disturbances, model error, or initial condition error. In this case, we can think of the controller \eqref{eqn:youla-ctrl} as consisting of a nominal controller $\mathcal{K}_b$ with good performance in response to known inputs, and a robustifying parameter $\mathcal{Q}$ whose role is to respond to error and uncertainty. This perspective is explored in the linear setting in \cite{Zhou+Ren2001}. 
\end{remark}

%
%
\section{Results for Simplified Settings} \label{sec:minitheory}

Our aim is to study the Youla controller $\mathcal{K}_\mathcal{Q}$ as a parameterization of policies for (a) nonlinear (b) partially-observed systems achieving (c) contracting and Lipschitz closed loops. In this section, we first consider simplified settings in which at most two of these three features are present at a time, with our results summarized in the top half of Table~\ref{tab:youla-theory}. In these settings, we find that a controller stabilizes the closed-loop system if and only if it can be written in the Youla form. Proofs of Theorems~\ref{thm:youla-lti} and \ref{thm:youla-state-feedback} are included as special cases of the results in Section~\ref{sec:theory}. The proof of Theorem~\ref{thm:youla-finite-gain} is omitted due to space constraints, but similar results appear in \cite{Imura+Yoshikawa1997,Fujimoto+Sugie2000}.

\subsection{Restricting to LTI Systems}\label{sec:theory-linear}

The following result from\cite{Wang++2022} is included here for completeness.
Consider the simplified setting in which \eqref{eqn:system} is an LTI system
\begin{equation} \label{eqn:linear-system}
        x_{t+1} = A x_t + B u_t + w_t, \quad
        y_t = C x_t + v_t
\end{equation}
with $(A,B)$ stabilizable and $(A,C)$ detectable. Then, we can obtain a linear base controller $\hat{u}=\mathcal{K}_b(y)$ as
\begin{subequations}\label{eqn:linear-Kb}
    \begin{empheq}[left=\mathcal{K}_b:\;\empheqlbrace]{align}
    \hat{x}_{t+1} &=A\hat{x}_t+B u_t +  L(y_t-C\hat{x}_t) \label{eqn:Kb-a} \\
    \hat{u}_t &=-K\hat{x}_t
    \end{empheq}
\end{subequations}
where the gain matrices $K, L$ are chosen such that $(A - BK)$ and $(A - LC)$ are stable. By taking the innovation $\tilde{y}=y-C\hat{x}$, the Youla controller \eqref{eqn:youla-ctrl} becomes
\begin{equation} \label{eqn:youla-linear}
    u=-K\hat{x} + \tilde{u}
\end{equation} 
with $\tilde{u}=\mathcal{Q}(\eta,\tilde{y})$, where $\mathcal{Q}$ is the nonlinear system \eqref{eqn:youla-param}.

If $\mathcal{Q}$ is an LTI system, it is well-known that \eqref{eqn:youla-linear} parameterizes all linear stabilizing controllers for the plant \eqref{eqn:linear-system} if and only if $\mathcal{Q}$ itself is stable \cite{Doyle1984,Anderson1998}. We extended this classic result in \cite{Wang++2022} to a nonlinear $\mathcal{Q}$ of the form \eqref{eqn:youla-param} and nonlinear controllers $u = \mathcal{K}(\eta, y)$ with state-space realizations 
\begin{equation} \label{eqn:generic-ctrl}
    \mathcal{K}:\; \left\{\begin{aligned}
        \phi_{t+1} &= f_\mathcal{K}(\phi_t, \eta_t, y_t) \\
        u_t &= h_\mathcal{K}(\phi_t, \eta_t, y_t) \\
    \end{aligned}\right.
\end{equation}
states $\phi\in\mathbb{R}^{n_k}$, and $f_\mathcal{K}, h_\mathcal{K}$ locally Lipschitz.

\begin{theorem}[Prop.~1 from \cite{Wang++2022}] \label{thm:youla-lti}
    Consider the LTI system \eqref{eqn:linear-system} with controller \eqref{eqn:Kb-a}, \eqref{eqn:youla-linear} parameterized by $\mathcal{Q}$ \eqref{eqn:youla-param}.
    \begin{enumerate}
        \item For any contracting and Lipschitz $\mathcal{Q}$, the closed-loop system is contracting and the map $(\eta,d) \mapsto z$ is Lipschitz.
        \item Any controller $u=\mathcal{K}(\eta,y)$ of the form \eqref{eqn:generic-ctrl} achieving a contracting and Lipschitz closed loop can be written in the form \eqref{eqn:Kb-a}, \eqref{eqn:youla-linear} with contracting and Lipschitz $\mathcal{Q}$.
    \end{enumerate}
\end{theorem}

The Youla parameterization represents all and only contracting and Lipschitz closed loops for a given LTI system.

\subsection{Restricting to Perfect State Feedback} \label{sec:theory-state-feedback}

Next, we consider settings in which the plant \eqref{eqn:system} is nonlinear but we have perfect knowledge of the states such that
\begin{equation} \label{eqn:system-state-feedback}
    x_{t+1} = f(x_t, \eta_t, u_t) + w_t, \quad y_t = x_t
\end{equation}
where $f(x_{-1}, \eta_{-1}, u_{-1}) := 0$ and $w_{-1} := x_0$. Suppose the base controller \eqref{eqn:base-controller} takes the form $\hat{u}_t = k(\eta_t, y_t)$.
Then the Youla parameterization \eqref{eqn:youla-ctrl} can be reduced to 
\begin{equation} \label{eqn:youla-state-feedback}
    u = k(\eta, x) + \tilde{u}
\end{equation}
with $\tilde{u}=\mathcal{Q}(\eta,\tilde{y})$, where the innovations at time $t$ are simply
\begin{equation} \label{eqn:state-feedback-innovations}
    \tilde{y}_t=w_{t-1}=x_t-f(x_{t-1},\eta_{t-1},u_{t-1})
\end{equation}
since we have a perfect model of the plant and no measurement noise. A similar statement to Theorem~\ref{thm:youla-lti} is then as follows.

\begin{theorem} \label{thm:youla-state-feedback}
    Suppose that $k$ is Lipschitz, the closed-loop system  under the base controller $\hat{u} = k(\eta, x)$ is contracting, and the map $(\eta, w, \tilde{u}) \mapsto z$ is Lipschitz.
    Consider the system \eqref{eqn:system-state-feedback} with controller \eqref{eqn:youla-state-feedback} parameterized by $\mathcal{Q}$ \eqref{eqn:youla-param}.
    \begin{enumerate}
        \item For any contracting and Lipschitz $\mathcal{Q}$, the closed-loop system is contracting and its map $(\eta, d) \mapsto z$ is Lipschitz.
        \item Any controller $u=\mathcal{K}(\eta, y)$ of the form \eqref{eqn:generic-ctrl} achieving a contracting and Lipschitz closed-loop can be written in the form \eqref{eqn:youla-state-feedback} with contracting and Lipschitz $\mathcal{Q}$.
    \end{enumerate}
\end{theorem}

As with the LTI setting, the Youla parameterization represents all and only contracting and Lipschitz closed-loops with \eqref{eqn:system-state-feedback}.
This formulation has strong connections to disturbance-feedback parameterizations in MPC \cite{Goulart++2006}. Similar results to Theorem~\ref{thm:youla-state-feedback} in terms of finite-gain stability have also been derived for system-level synthesis (SLS) \cite{Ho2020,Furieri++2022}.

\subsection{Restricting to Finite-Gain Stability} \label{sec:theory-finite-gain}

Moving to the partially-observed nonlinear setting, similar results have been proven for the weaker notion of finite-gain stability (e.g., \cite{Imura+Yoshikawa1997,Fujimoto+Sugie2000}). We summarize these results for our specific problem setup given the following assumptions.
\begin{enumerate}[label=\textbf{A\arabic*}]   
    \item[\textbf{A1}$^\prime$] \customlabel{\textbf{A1}$^\prime$}{assump:stabilizing-base} \textit{Finite-gain stabilizing base controller:} the closed-loop system $\mathcal{G}_{\mathcal{K}_b}: (\eta,d,\tilde{u})\mapsto z$ under the base controller, i.e., \eqref{eqn:system} in feedback with \eqref{eqn:base-controller}, is finite-gain stable. 

    \setcounter{enumi}{1}

    \item \textit{Contracting \& Lipschitz observer:} The observer \eqref{eqn:observer} is contracting and the map $(\eta, u,y) \mapsto \hat{x}$ is Lipschitz. \label{assump:observer-contract}

    \item \textit{Observer correctness:} When $d \equiv 0$ and $x_0 = \hat{x}_0$, the observer exactly replicates the plant dynamics. That is, $f(x_t, \eta_t, u_t) = f_o(x_t, \eta_t, u_t, h(x_t)) \ \forall t\in \mathbb{N}$. \label{assump:observer-correctness}

    \item \textit{Smooth functions:} The systems $\mathcal{G}, \mathcal{K}_b, \mathcal{O}, \mathcal{Q}, \mathcal{K}$ have piecewise-differentiable dynamics functions and Lipschitz continuous output maps.
    \label{assump:smooth-functions}
\end{enumerate}

Comparing to the partially-observed LTI setting, assumptions~\ref{assump:stabilizing-base}--\ref{assump:observer-correctness} are similar to assuming that we can design a stabilizing linear controller \eqref{eqn:linear-Kb}. Importantly, \ref{assump:observer-contract}--\ref{assump:observer-correctness} imply that the observer error exponentially decays when there are no disturbances, regardless of the control inputs \cite[Prop.~2]{Revay++2023}. They do not imply that the plant is contracting or Lipschitz---even in the linear regime, one can easily construct unstable systems with stable (contracting and Lipschitz) and correct observers that perfectly track the plant state when $d\equiv 0$. Under these assumptions, the Youla parameterization achieves the following result (similar to \cite[Prop.~21]{Fujimoto+Sugie2000}, \cite[Thm.~3.1]{Imura+Yoshikawa1997}).

\begin{theorem} \label{thm:youla-finite-gain}
    Suppose assumptions \ref{assump:stabilizing-base}--\ref{assump:smooth-functions} hold. Consider the system \eqref{eqn:system} with controller \eqref{eqn:youla-ctrl} parameterized by $\mathcal{Q}$ \eqref{eqn:youla-param}.
    \begin{enumerate}
        \item For any finite-gain stable $\mathcal{Q}$, the closed-loop system is also finite-gain stable.
        \item If the base controller \eqref{eqn:base-controller} is finite-gain stable, then any controller $u=\mathcal{K}(\eta, y)$ of the form \eqref{eqn:generic-ctrl} achieving a finite-gain stable closed loop can be written in the form \eqref{eqn:youla-ctrl} with finite-gain stable $\mathcal{Q}$.
    \end{enumerate}
\end{theorem}

This implies that the Youla parameterization represents all and only finite-gain stabilizing controllers for a given partially-observed nonlinear system.

%
%
\section{Results for Partially-Observed Nonlinear Systems with Incremental Stability} \label{sec:theory}

We now study the case of partially-observed nonlinear systems \eqref{eqn:system} and the incremental stability properties of contraction and Lipschitzness. We first show by counterexample that closed-loop incremental stability can be lost in this context. We follow by introducing a weaker form of stability which is maintained, and discuss stronger conditions under which contracting and Lipschitz closed loops can still be achieved.
Before continuing, we introduce a stronger version of Assumption~\ref{assump:stabilizing-base} which will be used for all results in this section
\begin{enumerate}[label=\textbf{A\arabic*}]
    \item \textit{Contracting \& Lipschitz stabilizing base controller:} the closed-loop system $\mathcal{G}_{\mathcal{K}_b}: (\eta,d,\tilde{u})\mapsto z$ under the base controller, i.e., \eqref{eqn:system} in feedback with \eqref{eqn:base-controller}, is contracting and Lipschitz. \label{assump:contracting-base}
\end{enumerate}
All proofs are provided in Appendices~\ref{app:proof-main-result}--\ref{app:proof-iqc}.

\subsection{Counterexample Exhibiting Loss of Contraction} \label{sec:theory-nonlinear}

One might expect a similar result to Theorems~\ref{thm:youla-lti}--\ref{thm:youla-finite-gain} to hold for partially-observed nonlinear systems in the incremental stability setting -- i.e.,
if the base controller satisfies assumptions \ref{assump:contracting-base}--\ref{assump:smooth-functions} and the Youla parameter is contracting and Lipschitz, then the Youla controller \eqref{eqn:youla-ctrl} parameterizes all contracting and Lipschitz closed loops $(\eta, d) \mapsto z$ for partially-observed nonlinear systems \eqref{eqn:system}. However, Example~\ref{ex:counter-example-1} shows a system for which these assumptions are satisfied, but the closed-loop system is neither contracting nor Lipschitz.

\begin{figure}[!t]
    \centering
    \includegraphics[trim={0.4cm 0.25cm 0cm 0.3cm},clip,width=0.9\columnwidth]{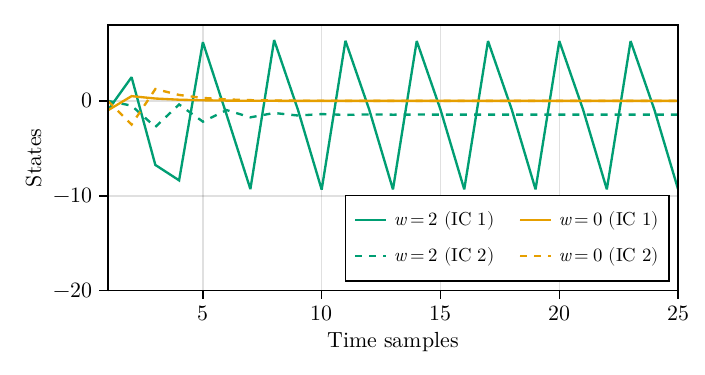}
    \caption{Simulations of the closed-loop system in Example~\ref{ex:counter-example-1} show plant state trajectories $x$ from initial conditions (ICs) $(x_0, \hat{x}_0) = (-1, 0), \ (0, -1)$. Observer states $\hat{x}$ are similar to $x$ and are omitted for neatness. The trajectories converge when $w = 0$. If $w=2$, there are distinct trajectories that do not converge, so the system is neither contracting nor Lipschitz.}
    \label{fig:counter-example-state}
\end{figure}

\begin{example}\label{ex:counter-example-1}
    Consider the following scalar nonlinear system
    \begin{equation}\label{eq:example-1-sys}
        x_{t+1}=0.5\abs{x_t}+(u_t-\eta_t)+w_t,\quad y_t=x_t
    \end{equation}
    with $w_t$ as an unknown disturbance and $\eta_t$ a known reference signal. Assumption \ref{assump:contracting-base} holds with the base controller $\mathcal{K}_b=0$. We then choose an observer 
    \begin{equation}\label{eq:example-1-obs}
        \hat{x}_{t+1}=0.5\abs{\hat{x}_t}+(u_t -\eta_t),\quad \hat{y}_t=\hat{x}_t,
    \end{equation}
    which satisfies assumptions \ref{assump:observer-contract}--\ref{assump:smooth-functions}. Finally, we pass the innovations $\tilde{y} = y - \hat{y}$ to a memoryless Youla parameter
    \begin{equation}\label{eq:example-Q}
        u=\mathcal{Q}(\tilde y):=\min(2.5-5\tilde y,0),
    \end{equation}
    which is obviously contracting and Lipschitz. As shown in Fig.~\ref{fig:counter-example-state} (green), the closed-loop states under the same disturbance $w \equiv 2$ converge to different trajectories depending on initial conditions, so the closed-loop system under the nonlinear Youla controller $\mathcal{K}_{\mathcal{Q}}$ is neither contracting nor Lipschitz. 
\end{example}

To understand the behavior of the system \eqref{eq:example-1-sys}--\eqref{eq:example-Q}, we first derive the dynamics of the observer error $\tilde{x}=x-\hat{x}$ as follows,
\begin{equation}\label{eq:obs-err-dyn}
    \tilde{x}_{t+1}=0.5|\tilde{x}_t+\hat{x}_t|-0.5|\hat{x}_t|+w_t,\quad \tilde y_t= \tilde x_t.
\end{equation}
The innovations $\tilde{y}$ implicitly depend on the control input $u$ (the output of $\mathcal{Q}$) through $\hat{x}$. This introduces a feedback loop between the error dynamics \eqref{eq:obs-err-dyn} and the Youla parameter $\mathcal{Q}$ \eqref{eq:example-Q}, leading to non-contracting closed-loop behavior when the gain of $\mathcal{Q}$ is sufficiently large. Note that if we remove the nonlinearity $|\cdot|$ to make the system linear, then the $\hat{x}_t$ terms in \eqref{eq:obs-err-dyn} cancel so that $\tilde{y}$ is independent of $u$, and the closed-loop system would be contracting and Lipschitz.

Despite the dependence of $\tilde{y}$ on $u$, we observed that increasing the gain of $\mathcal{Q}$ does not lead to unbounded closed-loop trajectories given bounded disturbances. The reason is as follows. Both the plant \eqref{eq:example-1-sys} and its observer \eqref{eq:example-1-obs} are copies of the following contracting and Lipschitz system 
\begin{equation}\label{eq:contract-lipschitz-sys}
    \xi_{t+1}=0.5\abs{\xi_t}+\mu_t, \quad \nu_t=\xi_t
\end{equation}
with states, inputs, and outputs $(\xi^a,\mu^a,\nu^a)=(x,u-\eta+w, y)$ and $(\xi^b,\mu^b,\nu^b)=(\hat{x},u-\eta,\hat{y})$ for the plant and observer, respectively. From \eqref{eqn:lipschitz} we have
\begin{equation}
    \|\tilde{y}\|_T = \|y - \hat{y}\|_T \leq \gamma \|w\|_T+\kappa(x_0,\hat{x}_0)
\end{equation}
with $\gamma>0$ and $\kappa(x_0,\hat{x}_0)\geq 0$. In other words, the innovations $\tilde y$ are bounded by the external disturbance $w$, and this bound is independent of $u$ despite there being feedback through $u$ which may be large due to a high-gain $\mathcal{Q}$. Moreover, since the Youla parameter $\mathcal{Q}$ and the system under the base controller are both contracting and Lipschitz, we can conclude that the $\ell_2$-norms of $u,x,y$ are also bounded, as expected from Theorem~3.1.

\subsection{d-Tube Contracting and Lipschitz Responses}\label{sec:d-tube}

The counterexample above shows that closed-loop contraction and Lipschitzness are not be guaranteed for partially-observed nonlinear systems \eqref{eqn:system}. However, we observed that the closed-loop state remained within a bounded tube whose size depended on the unknown disturbance, not the known inputs. We formalize this property with the following definition.

\begin{definition}[d-tube contracting and Lipschitz] \label{dfn:dtube}
    Consider the nonlinear system 
    \begin{equation}\label{eqn:sys}
        \chi_{t+1} = f(\chi_t,\eta_t,d_t), \quad z_t = h(\chi_t,\eta_t,d_t)
    \end{equation}
    with state $\chi$, output $z$, known input $\eta$, and unknown input $d$. The system \eqref{eqn:sys} is said to be \emph{d-tube contracting} if any pair of state trajectories $\chi^a,\chi^b$ with the same input $(\eta,d)$ and $\Delta \chi_t=\chi_t^a-\chi_t^b$ satisfies
    \begin{equation}\label{eqn:tube-contraction}
        |\Delta \chi_t|\leq  \kappa_1(\chi_0^a,\chi_0^b)\alpha^t+\gamma_1\|d\|_{t-1,\infty}\quad \forall t\in \mathbb{N}
    \end{equation}
    for some $\alpha\in [0,1)$, $\gamma_1\in \R^+$ and $\kappa_1(\chi_0^a,\chi_0^b)\geq 0$ with $\|d\|_{-1,\infty}:=0$. The system \eqref{eqn:sys} is \emph{d-tube Lipschitz} if for any pair of solutions $z^a, z^b$ with initial states $\chi_0^a, \chi_0^b$ we have
    \begin{equation}\label{eqn:tube-lipschitz}
        \begin{split}
            \norm{\Delta z}_T \le & 
                \gamma (\norm{\Delta \eta}_T + \norm{\Delta d}_T) + \gamma_2 
                (\norm{d^a}_T+\norm{d^b}_T) \\
                &+ \kappa_2(\chi_0^a, \chi^b_0)\quad \forall T\in \mathbb{N}
        \end{split}
    \end{equation}
    with $\gamma,\gamma_2\in \R^+$ and $\kappa_2(\chi_0^a,\chi_0^b)\geq 0$.
\end{definition}

The interpretation of Definition~\ref{dfn:dtube} is that the system responds smoothly to known inputs $\eta$ (e.g., disturbance previews, reference trajectories, feedforward controls), but the response to unknown inputs $d$ is only bounded (not necessarily smooth). We call it ``d-tube'' because if $\eta^a = \eta^b$, $d^a = d$, and $d^b = 0$, the response $(\chi^a,z^a)$ converges to a tube around the disturbance-free trajectory $(\chi^b,z^b)$ with radius proportional to the norm of $d$. This is a stronger property than finite-gain stability, but weaker than contraction and Lipschitzness in both $\eta$ and $d$. Another key difference from contraction and Lipschitzness is that we do not assume $\kappa_i(\chi_0, \chi_0) = 0$ in \eqref{eqn:tube-contraction} and \eqref{eqn:tube-lipschitz}, which allows for transient responses to initial conditions.

We now establish that d-tube contracting and Lipschitz closed-loops are guaranteed by the Youla parameterization in the partially-observed nonlinear setting. In the following, we use $\bar{x}_t := [x_t; s_t; q_t]$ to denote the state of the composition of the Youla parameter $\mathcal{Q}$ and the base-controlled system $\mathcal{G}_{\mathcal{K}_b}$, and $\tilde{x}_t := x_t - \hat{x}_t$ for the observer error. The total closed-loop state is $\chi_t=[\bar{x}_t; \tilde{x}_t]$.

\begin{theorem} \label{thm:youla-disturbance-tube}
    Consider the system \eqref{eqn:system} with controller \eqref{eqn:youla-ctrl} parameterized by the Youla parameter $\mathcal{Q}$ \eqref{eqn:youla-param}. Suppose assumptions \ref{assump:contracting-base}--\ref{assump:smooth-functions} hold and $f, f_b, f_q$ are Lipschitz continuous \textit{w.r.t.} their system inputs. Then, for any contracting and Lipschitz $\mathcal{Q}$, the closed-loop system is d-tube contracting and Lipschitz with 
    \begin{equation} \label{eqn:dtube-contraction-youla}
        \kappa_i(\chi_0^a,\chi_0^b)=\bar{\gamma}_i\abs{ \bar{x}_0^a-\bar{x}_0^b} + \tilde{\gamma}_i(|\tilde{x}_0^a| + |\tilde{x}_0^b|)
    \end{equation}
    for $\bar{\gamma}_{i},  \tilde{\gamma}_i\in\mathbb{R}^+$ and $i=1,2$.
\end{theorem}

The key property enabling this result is that despite the coupling between the innovations $\tilde{y}$ and the Youla augmentation $\tilde{u}$ which destroys closed loop contraction and Lipschitzness (Example~\ref{ex:counter-example-1}), the contraction and Lipschitz properties of the observer ensure that the coupling is bounded. Specifically, we show in the proof of Theorem~\ref{thm:youla-disturbance-tube} (Appendix~\ref{app:proof-main-result}) that the observer error $\tilde{x}$ (and hence also the innovations $\tilde{y}$) is constrained within a tube whose size is independent of $\tilde{u}$ with
\begin{equation}\label{eqn:obsv-err-bound}
    \begin{split}
        |\tilde{x}_t| &\le  \beta_o \alpha_o^t |\tilde{x}_0|+\frac{\beta_o \sqrt{\gamma_o^2+1}}{1-\alpha_o}\|d\|_{t-1,\infty} 
    \end{split}
\end{equation}
where $\alpha_o, \beta_o,$ and $\gamma_o$ are the contraction rate, overshoot, and Lipschitz bound of the observer, respectively.
Using this bound allows us to obtain finite gain bounds for the closed-loop response w.r.t. $d$ as the Youla parameter $\mathcal{Q}$ and $\mathcal{G}_{\mathcal{K}_b}$ are both contracting and Lipschitz by assumption.

An interesting special case of Theorem~\ref{thm:youla-disturbance-tube} arises when there is no uncertainty in the system. In this case, the closed-loop response reverts to being contracting and Lipschitz in the sense of Definitions~\ref{dfn:contraction} and \ref{dfn:lipschitz}. We formalize this as follows.
\begin{corollary} \label{cor:youla-cl}
    If there is no uncertainty ($d= 0$, $\tilde{x}_0=0$), then the closed-loop interconnection of \eqref{eqn:system} and \eqref{eqn:youla-ctrl} is contracting and $\eta \mapsto z$ is Lipschitz.
\end{corollary}

\begin{remark}
    The closed-loop system responds differently to known and unknown inputs $\eta$ and $d$ because $\eta$ is seen by the observer. Since the observer is itself contracting and Lipschitz, it responds smoothly to changes in $\eta$, and propagates this behavior through the closed loop via the innovations $\tilde{y}$.
\end{remark}


Another interesting special case of Theorem~\ref{thm:youla-disturbance-tube} arises when the only unknown quantity is the initial plant state -- i.e., $d = 0$ but the initial observer error is $\tilde{x}_0 \ne 0$. From \eqref{eqn:obsv-err-bound}, the observer error (and hence the innovations $\tilde{y}$) is constrained within an envelope that decays exponentially when $d = 0$.
Since $\mathcal{Q}$ and $\mathcal{G}_{\mathcal{K}_b}$ are both contracting and Lipschitz, the closed-loop state smoothly exponentially converges to some trajectory, which we see empirically in Fig.~\ref{fig:counter-example-state} (yellow), and theoretically by combining \eqref{eqn:tube-contraction} with \eqref{eqn:dtube-contraction-youla} when $d = 0$:
\begin{equation}
    \begin{aligned}
    \abs{\Delta \chi_t} &\le \bigl(\bar{\gamma}_1  \abs{\Delta \bar{x}_0} + \tilde{\gamma}_1(|\tilde{x}_0^a| + |\tilde{x}_0^b|)\bigr)\alpha^t 
    \end{aligned}
\end{equation}
Although $\abs{\Delta \chi_t}$ decays exponentially, the closed loop is not contracting in the sense of Definition \ref{dfn:contraction} since $\kappa_1(\chi_0^a,\chi_0^b)=\tilde{\gamma}_1(|\tilde{x}_0^a| + |\tilde{x}_0^b|)\neq 0$ even when $\Delta \bar{x}_0=0$ and $\Delta \tilde{x}_0=0$. These terms are due to the transient feedback interaction between $\tilde{u}$ and $\tilde{y}$ before the observer error $\tilde{x}$ converges to 0. We called this behavior contracting and Lipschitz \emph{with transients} in \cite{Barbara++2023a}, and it is consistent with recent results in \cite[Thm.~4.3, 4.4]{Kawano++2024}.

\subsection{A Unifying Sufficient Condition when Innovations are Decoupled from the Controller Augmentation}

The source of the loss of contraction in the counterexample (Sec. \ref{sec:theory-nonlinear}) was the  interaction between the augmentation input $\tilde u$ and the innovations $\tilde y$. If, however, $\tilde{y}$ is completely decoupled from $\tilde{u}$, then adding $\mathcal{Q}$ does not introduce a feedback loop, and we retain the closed-loop contracting and Lipschitz properties of the original base-controlled system.
We have already seen this property in three special cases -- LTI systems (Thm.~\ref{thm:youla-lti}.1), fully-observed nonlinear systems (Thm.~\ref{thm:youla-state-feedback}.1), and partially-observed nonlinear systems with zero uncertainty (Cor.~\ref{cor:youla-cl}). 

\begin{proposition} \label{prop:innov-system}
    In the closed-loop systems from Theorems\,\ref{thm:youla-lti},\,\ref{thm:youla-state-feedback}, and Corollary\,\ref{cor:youla-cl}, the innovations $\tilde{y}$ are the outputs of a contracting and Lipschitz system $\mathcal{T}: (\eta, d) \mapsto \tilde{y}$ with
    \begin{equation} \label{eqn:innovations-system}
        \mathcal{T}:\; \left\{\begin{aligned}
            \psi_{t+1} &= \tilde{F}(\psi_t, \eta_t, d_t) \\
            \tilde{y}_t &= \tilde{H}(\psi_t, \eta_t, d_t)
        \end{aligned}\right.,
    \end{equation}
    internal state $\psi_t \in \mathbb{R}^{n_\psi}$, and $\tilde{F}, \tilde{H}$ locally Lipschitz.
\end{proposition}

The decoupling of $\tilde{y}$ and $\tilde{u}$ in this way is a sufficient condition for the Youla parameterization to guarantee contracting and Lipschitz closed-loop responses. We therefore unify our three special cases with the following result.

\begin{theorem} \label{thm:youla-decoupled}
    Suppose the base-controlled system $\mathcal{G}_{\mathcal{K}_b}: (\eta,d,\tilde{u})\mapsto z$ satisfies assumption \ref{assump:contracting-base} and the innovations $\tilde{y}$ are generated by a contracting and Lipschitz system $\mathcal{T}$ \eqref{eqn:innovations-system}. Then for any contracting and Lipschitz Youla parameter $\mathcal{Q}$ \eqref{eqn:youla-param}, the interconnected system $z = \mathcal{G}_{\mathcal{K}_b} ( \eta, d, \mathcal{Q}( \eta, \mathcal{T}(\eta, d)))$ mapping $(\eta, d) \mapsto z$ is also contracting and Lipschitz.
\end{theorem}

\subsection{Sufficient Condition for the Coupled Case and Uncertain Models via Frequency-Weighted Gain Bounds } \label{sec:theory-uncertain}

\begin{figure}[!tb]
    \centering
    \includegraphics[width=0.7\linewidth]{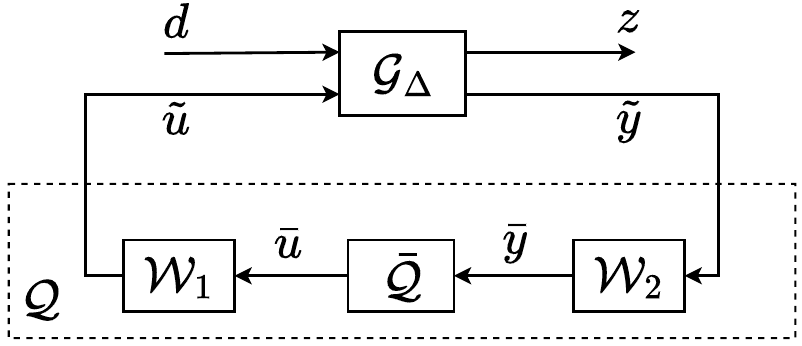}
    \caption{Youla parameterization via loop shaping, where known inputs $\eta$ have been omitted to simplify the diagram.}
    \label{fig:loop-shaping}
\end{figure}

When the innovations are coupled with the Youla augmentation (e.g., due to disturbances or model parameter uncertainty), we can still achieve contracting and Lipschitz closed loops by imposing further restrictions on the Youla parameter.

Our approach is based on IQC theory and loop shaping. We first group together the nonlinear plant $\mathcal{G}$, the base controller $\mathcal{K}_b$, and the observer $\mathcal{O}$ as the system $\mathcal{G}_{\Delta}:(\eta, d,\tilde{u})\mapsto (z,\tilde{y}$) shown in Fig.~\ref{fig:loop-shaping}. We then take $\mathcal{Q}=\mathcal{W}_1 \circ \bar{\mathcal{Q}} \circ \mathcal{W}_2$ where $\mathcal{W}_1$ and $\mathcal{W}_2$ are fixed, stable linear filters, and $\bar{\mathcal{Q}}$ is a learnable contracting and Lipschitz system. $\mathcal{Q}$ is therefore also contracting and Lipschitz. The following proposition gives a sufficient condition for achieving a stable closed loop.

\begin{proposition} \label{prop:iqc}
    Suppose the system $\mathcal{G}_{\Delta}$ is contracting and there exist stable linear filters $\mathcal{W}_1:\bar{u}\mapsto \tilde{u}$ and $\mathcal{W}_2:\tilde{y}\mapsto \bar{y}$ with $\bar{u}\in \ell^{n_u}$ and $\bar{y}\in \ell^{n_y}$ such that the weighted system $\bar{\mathcal{G}}_\Delta:(\eta, d,\bar{u})\mapsto (z,\bar{y})$ admits the incremental IQC defined by $\sigma_{\bar{\mathcal{G}}}$. Take $\mathcal{Q}=\mathcal{W}_1 \circ \bar{\mathcal{Q}}\circ \mathcal{W}_2$, and suppose that the system $\bar{\mathcal{Q}}:\bar{y}\mapsto \bar{u}$ is contracting and satisfies the incremental IQC defined by $\sigma_{\bar{\mathcal{Q}}}$. The closed loop  of $\mathcal{G}_{\Delta}$ and $\mathcal{Q}$ is contracting and Lipschitz if $\sigma_{\bar{\mathcal{G}}}+\sigma_{\bar{\mathcal{Q}}}$ is negative definite \textit{w.r.t.} $(\Delta \bar{y},\Delta \bar{u},\Delta z)$.
\end{proposition}

\begin{remark} \label{rmk:frequency-iqc}
Using a pre-compensator $\mathcal{W}_1$ and/or a post-compensator $\mathcal{W}_2$, the system $\bar{\mathcal{G}}_{\Delta}$ has a desired open-loop shape, which can be understood as a dynamic version of input normalization in deep RL \cite{Andrychowicz++2020}.  
If a Lipschitz bound type IQC is used, then the above result is reduced to the incremental small-gain theorem, i.e., $\mu\gamma < 1$ where $\gamma, \mu$ are Lipschitz bounds for $\bar{\mathcal{G}}_\Delta$ and $\bar{\mathcal{Q}}$, respectively.
\end{remark}

%
%
\section{Converse Results} \label{sec:converse}

In Section~\ref{sec:minitheory} we saw that, in certain simplified settings, the proposed Youla parameterization includes all possible and only those controllers resulting in contracting and Lipschitz closed-loops. In Section \ref{sec:theory} we saw by counterexample that this is not true in the partially-observed nonlinear case, but that the weaker property of d-tube contraction and Lipschitzness holds.
This raises the question: what class of controllers \textit{are} covered by the Youla parameterization? We provide an answer in this section, and include proofs in Appendix~\ref{app:proof-partial-converse}.

\begin{theorem} \label{thm:partial-converse}
Suppose assumption \ref{assump:smooth-functions} holds and that (a) the base controller $\mathcal K_b: (\eta, u, y)\mapsto \hat u$ is contracting and Lipschitz, and (b) the  controller $\mathcal K: (\eta, y) \mapsto u$ stabilizes the observer in the sense that the feedback system
\begin{subequations} \label{eqn:innov-obsv-base-ctrl}
    \begin{align}
      \mathcal O:\; & \begin{cases}
          \hat{x}_{t+1} = f_o(\hat{x}_t, \eta_t, u_t, \hat{y}_t+ \tilde{y}_t) \\ 
          \hat{y}_t = h(\hat{x}_t)
      \end{cases}  \label{eqn:innov-obsv-base-ctrl-observer} \\ 
     \mathcal K:\; &\begin{cases}
         \phi_{t+1} = f_{\mathcal K}(\phi_t, \eta_t, \hat y_t+ \tilde{y}_t) \\
         u_t = h_{\mathcal K}(\phi_t,\eta_t, \hat y_t+\tilde{y}_t)
     \end{cases} 
    \end{align}
\end{subequations}
is contracting and the map $(\eta, \tilde{y})\mapsto \hat{z}$ is Lipschitz, where $\hat{z}_t := [\hat{x}_t; u_t]$. Then $\mathcal K$ can be represented in the form \eqref{eqn:youla-ctrl} with a contracting and Lipschitz $\mathcal{Q}$.
\end{theorem}

\begin{figure}[!t]
    \centering
    \includegraphics[width=0.75\linewidth]{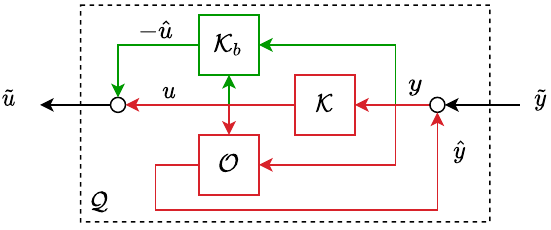}
    \caption{Block diagram showing how a Youla parameter $\tilde{u} = \mathcal{Q}(\tilde{y})$ can be constructed so that any controller $u = \mathcal{K}(y)$ can be realized in the Youla form \eqref{eqn:youla-ctrl}. Known exogenous inputs $\eta$ have been omitted to simplify the diagram.}
    \label{fig:youla-parameter}
\end{figure}

The main idea is that for a given base controller and observer, any controller $\mathcal{K}$ can be realized in the Youla form \eqref{eqn:youla-ctrl} with $\mathcal{Q}$ as the mapping $(\eta, \tilde y)\mapsto \tilde u$ shown in Fig.~\ref{fig:youla-parameter} -- i.e., a feedback interconnection of the controller $\mathcal K$ and the observer $\mathcal O$ (red), which is connected in series/parallel with the base controller $\mathcal K_b$ (green). With this construction, the mapping $y\mapsto u$ of the controller $\mathcal K$ remains unchanged. The conditions of the theorem imply that this $\mathcal Q$ is contracting and Lipschitz. We now offer some remarks.

\begin{remark}
Stabilizing the observer is related to stabilizing the plant \eqref{eqn:system} since when $\tilde{y} = 0$, observer correctness (\ref{assump:observer-correctness}) implies $f_o(x_t, \eta_t, u_t, h(x_t)) = f(x_t, \eta_t, u_t)$, which is the disturbance-free plant dynamics. The controller $\mathcal K$ stabilizes the observer against the particular disturbance structure
$$
\hat{x}_{t+1} = f_o(\hat{x}_t, \eta_t, u_t, h(x_t) + w_t)
$$
entering via the measurement channel, rather than simply an additive disturbance as in \eqref{eqn:system}.
\end{remark}

\begin{remark} Stability of the base controller may seem like a restrictive assumption, similar to the notion of strong stabilization  of linear systems. However, this is not the case since $\mathcal{K}_b$ takes the true control signal $u$ as an input. For example, this requirement is automatically satisfied if the base controller consists of a contracting and Lipschitz observer feeding a state estimate to a Lipschitz static state feedback. In the LTI case, this is the natural requirement that $(A - LC)$ is stable, whereas strong stabilization requires that $(A - LC - BK)$ be stable, which is restrictive. 
\end{remark}

\begin{remark}
    Stabilizing the red system in Fig.~\ref{fig:youla-parameter} can be interpreted as defining a stable image representation of the controller $\mathcal{K}$ (i.e., a right-coprime factor), since it represents a mapping from $\tilde{y} \mapsto (\mathcal{K}(y), y)$. Similarly, one can think of the  observer \eqref{eqn:observer} together with the innovations map \eqref{eqn:innovations} as a stable kernel representation (i.e., a left-coprime factor) mapping $(u, y) \mapsto \tilde{y}$. See also \cite{Fujimoto+Sugie2000,Kawano++2024}.
\end{remark}

In certain special cases, the requirement that  $\mathcal{K}$ stabilizes the observer follows naturally from $\mathcal{K}$ stabilizing the plant. Note that the converse part of Theorem~\ref{thm:youla-lti}.2 for LTI systems follows directly from Corollary~\ref{cor:partial-conv-certainty-equivalence} below.

\begin{corollary} \label{cor:partial-conv-special-cases} The converse part of
    Theorem~\ref{thm:youla-state-feedback}.2 follows  from Theorem~\ref{thm:partial-converse} with measurements $y_t = x_t$, observer $\hat{x}_{t+1} = f(y_t,\eta_t,u_t)$, and a static base controller $\hat{u}_t = k(\eta_t, y_t)$.
\end{corollary}

\begin{corollary}\label{cor:partial-conv-certainty-equivalence}
    Suppose assumption \ref{assump:smooth-functions} holds and the base controller \eqref{eqn:base-controller} and observer \eqref{eqn:observer} are in the certainty-equivalence form 
    \begin{equation} \label{eqn:observer-linear-innovations}
        \begin{aligned} 
            \hat{x}_{t+1} &= f(\hat{x}_t, \eta_t, u_t) + L(\eta_t, y_t - h(\hat{x}_t))  \\
            \hat{u}_t &= k(\hat{x}_t, \eta_t, y_t),
        \end{aligned}
    \end{equation}
    where $L: \mathbb{R}^{n_\eta} \times \mathbb{R}^{n_y} \rightarrow \mathbb{R}^{n_x}$ is Lipschitz. 
    Then any controller $u = \mathcal{K}(\eta, y)$ of the form \eqref{eqn:generic-ctrl} achieving a contracting and Lipschitz closed loop $(\eta,d)\mapsto z$ with \eqref{eqn:system} can be written in the form \eqref{eqn:youla-ctrl} with a contracting and Lipschitz $\mathcal{Q}$ \eqref{eqn:youla-param}.
\end{corollary}

\begin{figure}[!t]
    \centering
    \includegraphics[trim={7.5cm 5cm 9cm 4.8cm},clip,width=0.75\linewidth]{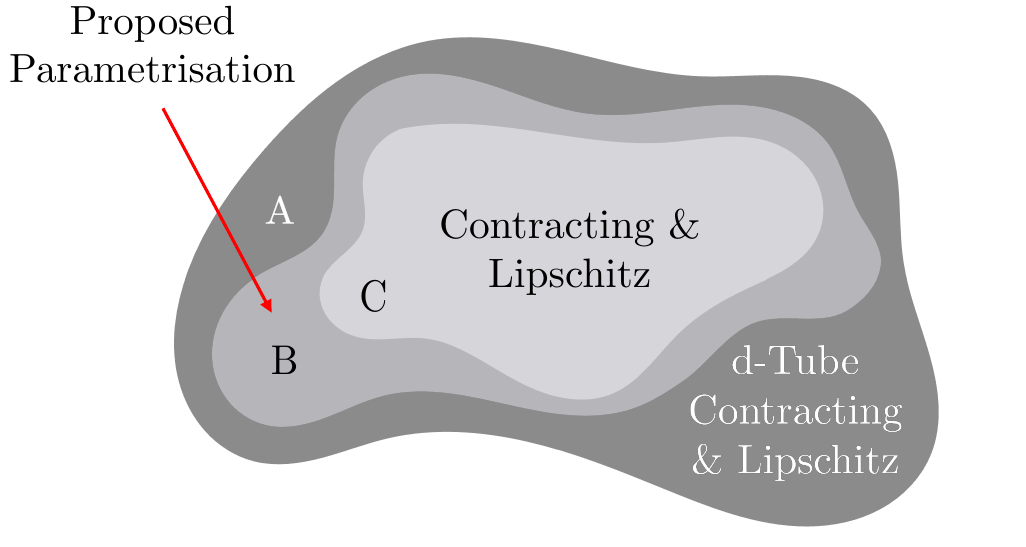}
    \caption{Illustration of stabilizing policy sets for controllers satisfying the conditions in Corollaries~\ref{cor:partial-conv-special-cases} and \ref{cor:partial-conv-certainty-equivalence}: (A) controllers achieving d-tube contracting and Lipschitz closed-loops, (B) the proposed nonlinear Youla parameterization, and (C) controllers achieving contracting and Lipschitz closed loops. The three sets coincide in the LTI and full-state feedback settings.}
    \label{fig:stabilizing-controller-sets}
\end{figure}

Fig.~\ref{fig:stabilizing-controller-sets} illustrates the connection between Theorems~\ref{thm:youla-disturbance-tube}, \ref{thm:youla-decoupled}, and the two corollaries of Theorem~\ref{thm:partial-converse}. For closed-loop systems satisfying both the structure and assumptions in Corollaries~\ref{cor:partial-conv-special-cases} or \ref{cor:partial-conv-certainty-equivalence}, the Youla parameterization \eqref{eqn:youla-ctrl} contains all policies \eqref{eqn:generic-ctrl} that achieve contracting and Lipschitz closed loops with the plant (set C). When the innovations and control inputs are coupled due to additive disturbances, the closed-loop system does not necessarily achieve these same strong stability properties. Instead, it guarantees d-tube contraction and Lipschitzness, which is a slightly weaker stability result (set A). In the special case where the innovations are decoupled from the control signal as in Theorem~\ref{thm:youla-decoupled}, all three sets coincide (e.g., LTI systems, full-state feedback).

%
%
\input{numerical_experiments}

%
%
\section{Conclusions} \label{sec:conc}

We have studied the problem of control policy parameterization for learning-based control, via a version of the Youla parameterization for nonlinear dynamical systems and a constructive neural network implementation called the Youla-REN. We have proven that: the parameterization guarantees contracting and Lipschitz closed loops when the innovations are decoupled from the Youla augmentation; if the signals are coupled due to additive disturbances, the closed-loop system responds smoothly to known inputs while disturbed trajectories are constrained within bounded tubes centered on the undisturbed trajectories; and we cover \textit{all} contracting and Lipschitz closed loops for certain systems. Using RENs as the Youla parameter, empirical studies showed that we can learn stabilizing and performant Youla-REN policies on feedback control tasks, even if they are trained with arbitrary (potentially de-stabilizing) cost functions, over short time horizons, and with model uncertainty. 

We conclude with key lessons and thoughts regarding practical application. Compared to black-box architectures, structured policies offer not only stability guarantees but also better empirical behavior both during training and in the loop long-term. The trade-off is the need for prior information.
Compared to residual RL, the Youla framework permits more aggressive reactions to disturbances without sacrificing stability. The price is the requirement for a good observer (and implicitly, a good system model), motivating future work on contracting observers for complex real-world systems.
The different behaviors of known and unknown external inputs ($\eta$ and $d$) suggest that disturbances should be estimated as accurately as possible, with the known component absorbed into $\eta$ to minimize what remains in $d$.
Finally, to guarantee stability, a critical quantity is the gain from the Youla augmentation to the innovations. Estimating tight bounds for complex systems (e.g., in robotics) will enable high-performance real-time learning with stability guarantees.

%
%

\appendix
\input{proofs}

\section*{References}
\bibliographystyle{IEEEtran}
\bibliography{references}

\vspace{-1cm}
\begin{IEEEbiography}[{\includegraphics[width=1in,height=1.25in,clip,keepaspectratio]{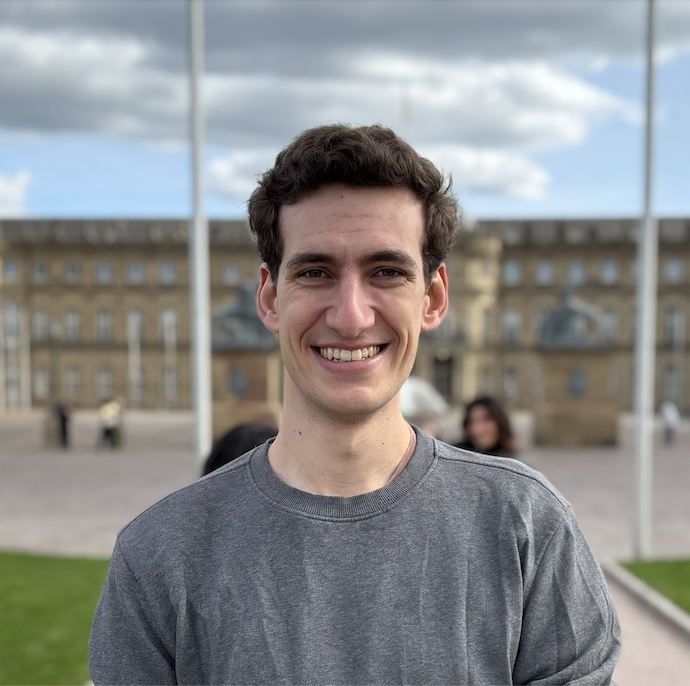}}]{Nicholas H. Barbara} 
    received the Ph.D. degree and the combined B.Sc. \& B.E. (Hons 1, University Medal) degrees from the University of Sydney, Australia, in 2025 and 2021 respectively. He is currently a postdoctoral research associate at the Centre for Complex Systems within The University of Sydney. His research interests include learning-based control, robust ML, time-series analysis, and their application to robotics and other complex systems.
\end{IEEEbiography}

\vspace{-1cm}
\begin{IEEEbiography}[{\includegraphics[width=1in,height=1.25in,clip,keepaspectratio]{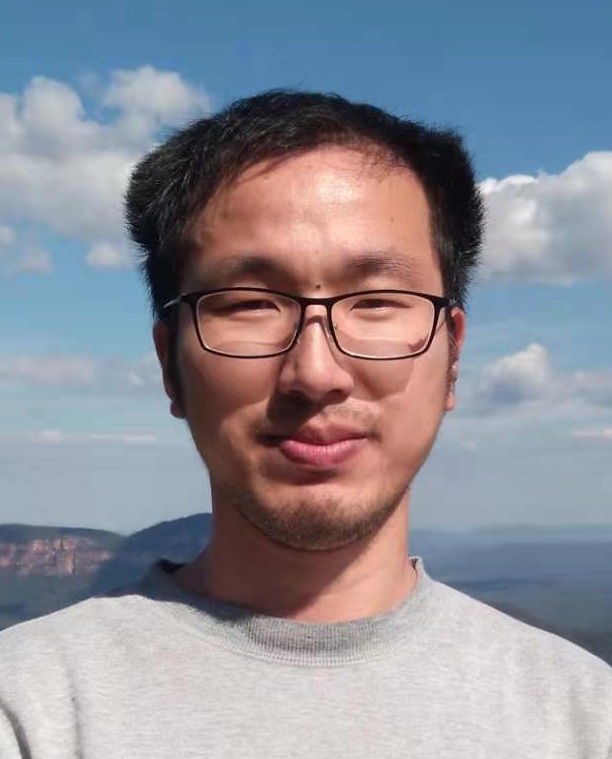}}]{Ruigang Wang} 
    received his Ph.D. degree in chemical engineering from The University of New South Wales (UNSW), Sydney, Australia, in 2017. From 2017 to 2018 he worked as a postdoc at UNSW. He is currently a postdoc at the Australian Centre for Robotics, The University of Sydney, Australia. His research interests include contraction based control, estimation, and learning for nonlinear systems.
\end{IEEEbiography}

\vspace{-1cm}
\begin{IEEEbiography}[{\includegraphics[width=1in,height=1.25in,clip,keepaspectratio]{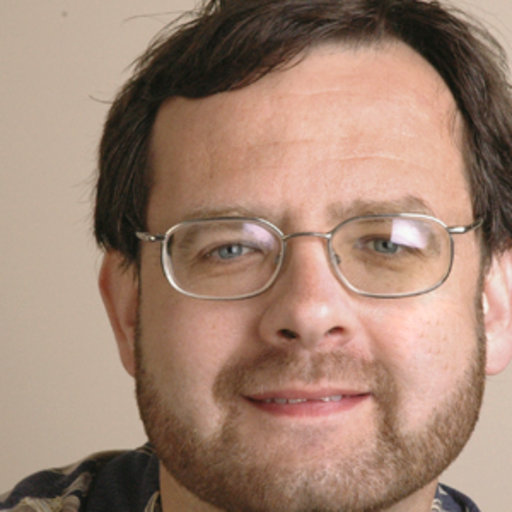}}]{Alexandre Megretski}
    received the graduate degree from Leningrad University, Russia, in 1988, and the Ph.D. degree in control theory. He held the post-doctoral positions with Mittag-Leffler Institute and the Royal Institute of Technology, Sweden, and the University of Newcastle, Australia. He was a Faculty Member with Iowa State University, and, since 1996, with Massachusetts Institute of Technology, where he is now a Professor of Electrical Engineering. His research interests include nonlinear dynamical systems, robust control, optimization, operator theory, model reduction, and system identification. He is also a co-founder of NanoSemi, Inc., Waltham, MA, USA, a company involved in digital compensation of analog circuits.
\end{IEEEbiography}
	
\vspace{-1cm}
\begin{IEEEbiography}[{\includegraphics[width=1in,height=1.25in,clip,keepaspectratio]{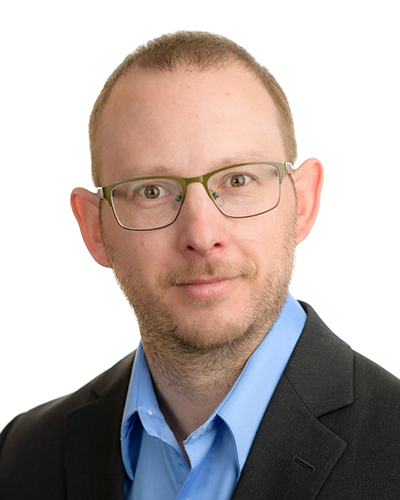}}]{Ian R. Manchester}
	received the B.E. (Hons 1) and Ph.D. degrees in Electrical Engineering from the University of New South Wales, Australia, in 2002 and 2006, respectively. He held research positions at Ume\aa\ University, Sweden, and Massachusetts Institute of Technology, USA. In 2012 he joined the faculty at the University of Sydney, where he is currently Professor of Mechatronic Engineering, Director of the Australian Centre for Robotics (ACFR) and the Australian Robotic Inspection and Asset Management Hub (ARIAM). His current research interests include robust machine learning and learning-based nonlinear control along with applications in robotics. He has served an Associate Editor for  IEEE Robotics \& Automation Letters and IEEE Control Systems Letters.
\end{IEEEbiography}

\end{document}

%% file: numerical_experiments.tex
\section{Deep RL with Youla-REN} \label{sec:exp}
Having established the stability properties of the nonlinear Youla parameterization \eqref{eqn:youla-ctrl}, we now investigate its use in learning-based control tasks. In this section, we first introduce our proposed Youla-REN policy class (Sec.~\ref{sec:youla-ren}), where we use a REN \cite{Revay++2023} as the contracting and Lipschitz Youla parameter $\mathcal{Q}$. We then illustrate the advantages of its built-in stability certificates on three deep RL tasks with: (i) general reward functions without stability consideration; (ii) short training horizons; and (iii) systems with model uncertainty (Sec.~\ref{sec:nl-example}--\ref{sec:exp-uncertainty}, respectively).
All experiments\footnote{\url{https://github.com/nic-barbara/ReactToSurprises}} were performed in Julia using the \texttt{RobustNeuralNetworks.jl} package \cite{Barbara++2025a} and the analytic policy gradients RL algorithm \cite{Freeman++2021,Wiedemann++2023}, which directly computes loss gradients for policy updates by backpropagating in time through the policy, system dynamics, and loss function with automatic differentiation.

\subsection{The Youla-REN Policy Class} \label{sec:youla-ren}

{We call the controller \eqref{eqn:youla-ctrl} a {\em Youla-REN} policy if the Youla parameter $\mathcal{Q}$ \eqref{eqn:youla-param} is a REN \cite{Revay++2023}, which is a nonlinear system
\begin{equation}\label{eq:ren}
    \begin{split}
        \begin{bmatrix}
            q_{t+1} \\ \bar{v}_t \\ \tilde{u}_t
        \end{bmatrix}=
        \overset{W}{\overbrace{
		\left[
            \begin{array}{c|cc}
            A & B_1 & B_2 \\ \hline 
            C_{1} & D_{11} & D_{12} \\
            C_{2} & D_{21} & D_{22}
		\end{array} 
		\right]
        }}
        \begin{bmatrix}
            q_t \\ \bar{w}_t \\ \tilde{y}_t
        \end{bmatrix}+
        \overset{b}{\overbrace{
            \begin{bmatrix}
                b_q \\ b_v \\ b_y
            \end{bmatrix}
        }}\\
        \bar{w}_t=\sigma(\bar{v}_t):=
        \begin{bmatrix}
            \sigma(\bar{v}_{t}^1) & \sigma(\bar{v}_{t}^2) & \cdots & \sigma(\bar{v}_{t}^{n_v})
        \end{bmatrix}^\top
    \end{split}
\end{equation} 
with learnable parameters $W, b$, where $ q_t \in \R^{n_q}$ is the internal state, and $ \bar{v}_t, \bar{w}_t\in \R^{n_{\bar{v}}}$ are the input and output of the neural layer, respectively. Here $ \sigma$ is a scalar activation with its slope restricted in $[0,1]$ (e.g., \texttt{relu} or \texttt{tanh}). The motivations for choosing a REN as the Youla parameter are as follows. 

\begin{enumerate}
    \item RENs admit built-in behavioral guarantees such as contraction, Lipschitzness, or other properties described by incremental IQCs. Thus, they are compatible with the analysis framework for our policy parameterization. In the experiments below, we use \emph{$\gamma$}REN to denote a REN with a prescribed Lipschitz bound of $\gamma\in\mathbb{R}^+$.
    \item RENs are highly flexible and include many existing models as special cases, including multi-layer perceptrons, recurrent neural networks, and stable linear dynamical systems. It was proven in \cite[Prop.~2]{Wang++2022} that contracting RENs are also universal approximators for contracting and Lipschitz dynamical systems.
    \item RENs are compatible with standard machine-learning tools as they permit a direct (smooth, unconstrained) parameterization $\theta\mapsto (W, b)$ such that for any $\theta \in \R^N$, the resulting REN $\mathcal{Q}_{\theta}$ \eqref{eq:ren} is contracting and Lipschitz. This enables training of large-scale models via automatic differentiation and gradient methods -- i.e., special optimization tools are not required to train a REN.
\end{enumerate}
Any contracting and Lipschitz model could be used as the Youla parameter $\mathcal{Q}$ in \eqref{eqn:youla-ctrl}. Note that RENs have two main limitations: solving an equilibrium layer at each iteration can be slow for large models, and the direct parameterization \cite{Revay++2023} makes it difficult to incorporate sparse structures (e.g., convolutions). Neither limitation is significant for our experiments, and concurrent work in \cite{Barbara++2025b} seeks to address both. See \cite{Manchester++2026} for a detailed review on  parameterizations of stable neural models including RENs.

\subsection{Nonlinear System with Economic Cost} \label{sec:nl-example}

In our first example, we learn stabilizing policies for partially-observed nonlinear system while optimizing an economic cost that does not naturally encourage stability. Consider the following academic example, adapted from \cite{andrieu2010uniting}:
\begin{equation} \label{eqn:nl-example-sys}
    \begin{aligned}
        \dot{x}_1 &= -x_1 + x_3 + w_1\\
        \dot{x}_2 &= x_1^2 - x_2 - 2x_1 x_3 + x_3 + w_2\\
        \dot{x}_3 &= -x_2 + u + w_3\\
        y &= [x_2; x_3] + v
    \end{aligned}
\end{equation}
with $w,v$ as zero-mean Gaussian noise with variances $\sigma_{w_i} = 10^{-2}, \sigma_v = 10^{-3}$ respectively.
We chose a stabilizing base controller $u = y_1 - ky_2$ with gain $k = 1.5$, and an observer of the form
\begin{equation} \label{eqn:nl-example-obs}
    \begin{aligned}
        \dot{\hat{x}}_1 &= -\hat{x}_1 + \hat{x}_3 \\
        \dot{\hat{x}}_2 &= \hat{x}_1^2 - \hat{x}_2 - 2\hat{x}_1 \hat{x}_3 + \hat{x}_3 \\
        \dot{\hat{x}}_3 &= -\hat{x}_2 + u + (y_2 - \hat{x}_3) - (y_1 - \hat{x}_2).
    \end{aligned}
\end{equation}
It can be shown that Assumptions \ref{assump:contracting-base}--\ref{assump:smooth-functions} hold for this base controller and observer via \cite{Wang++2024}. We discretized the system using the forward-Euler method with time-step of $\Delta t = 0.05$\,s.

\begin{figure}[!t]
    \centering
    \begin{subfigure}[b]{\linewidth}
        \centering
        \hspace{-1cm}
        \includegraphics[trim={0.2cm 7cm 0cm 0cm},clip,width=0.85\linewidth]{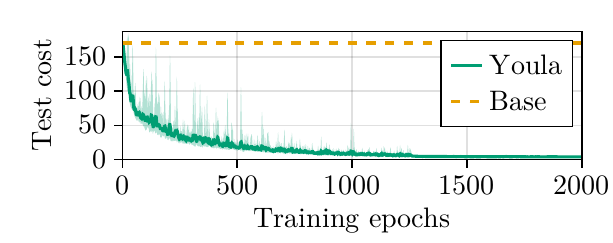}
        \caption{Test cost during training.}
        \label{fig:nl-traincurve}
    \end{subfigure}
    \begin{subfigure}[b]{\linewidth}
        \centering
        \hspace{-1cm}
        \includegraphics[trim={0.2cm 0.4cm 0cm 4cm},clip,width=0.85\linewidth]{Images/nonlinear_system_results_singlecol.pdf}
        \caption{Trajectory rollouts of $u$ and $x_2$ for a trained Youla-REN.}
        \label{fig:nl-rollouts}
    \end{subfigure}
    \caption{Youla-REN learns to improve economic performance while maintaining closed-loop stability. Bands in (a) show the range across 10 random seeds. Dotted lines in (b) indicate the soft constraints $\pm u_\mathrm{max}$ imposed on the controls.}
    \label{fig:nl-example}
\end{figure}

We chose an economic stage cost
\begin{equation} \label{eqn:nl-example-cost}
g(x_t, u_t) := |u_t| + 500 \max(|u_t| - u_\mathrm{max}, 0),
\end{equation}
which is the $\ell_1$-norm of the control inputs plus a strong penalty for exceeding the soft threshold $u_\mathrm{max}$. Note that if stability is ignored, then the optimal policy is simply $u\equiv 0$, which results in an unstable closed loop.

Fig.~\ref{fig:nl-example} shows that with the proposed Youla-REN, we are able to learn a stabilizing controller which achieves significant performance improvements compared to the base controller while maintaining closed-loop stability. The learned Youla parameter $\mathcal{Q}$ adds corrections to the base controller signal to keep the total $u$ within the soft bounds $\pm u_\mathrm{max}$ most of the time, allowing only small deviations outside the bounds for short intervals, and encouraging it close to zero within these bounds. Moreover, the trajectories of both $u$ and $x_2$ indicate that the Youla-REN preserves the closed-loop stability of the base controller even though such behavior was not explicitly encoded in the cost function. We therefore avoid the laborious process of reward-shaping to encourage stability in this task, which is commonly required for deep RL policies to be both stabilizing and performant (e.g., \cite{Rudin++2021}).

\subsection{Long-Term Stability with Short-Term Training} \label{sec:exp-doyle}

Our second example illustrates that our proposed Youla-REN policies can achieve good long-term test performance even when they are trained over short time horizons. We consider the following well-known linear quadratic Gaussian (LQG) example from \cite{Doyle1978},
\begin{equation} \label{eqn:doyle}
    \begin{aligned}
        \begin{bmatrix}
            \dot{x}_1 \\ \dot{x}_2
        \end{bmatrix} &= \mqty[1 & 1 \\ 0 & 1]\begin{bmatrix}
            x_1 \\ x_2
        \end{bmatrix} + \mqty[0 \\ 1] u + \mqty[1 \\ 1] w \\
        y &= x_1 + v,
    \end{aligned}
\end{equation}
where $w, v \in \mathbb{R}$ are Gaussian noises with variances $\sigma_w > 0$ and $\sigma_v = 1$, respectively. The control objective is 
\begin{equation} \label{eqn:lqg-costfunc}
    J = \mathbb{E}_{x_0,w, v} \int_{t=0}^\infty \left[q^2(x_1+x_2)^2 + u^2\right]\, dt 
\end{equation}
with $q > 0$. The optimal LQG controller has arbitrarily small-gain and phase margins as $\sigma_w, q $ increase \cite{Doyle1978}. This makes the problem challenging in a deep RL setting, since optimal and unstable policies may appear very close in parameter space.

We discretized the system with the forward-Euler method ($\Delta t = 0.01$\,s) and chose a training horizon of 100\,s. For the base controller, we chose a de-tuned LQG controller designed to minimize \eqref{eqn:lqg-costfunc} with $q = 10^3$ and $\sigma_w = 10$, while for training and test evaluations we used $q, \sigma_w = 10^3$. 

\begin{figure}[!t]
    \centering
    \begin{subfigure}[b]{\linewidth}
        \centering
        \includegraphics[trim={0.2cm 10.8cm 0cm 0cm},clip,width=0.75\linewidth]{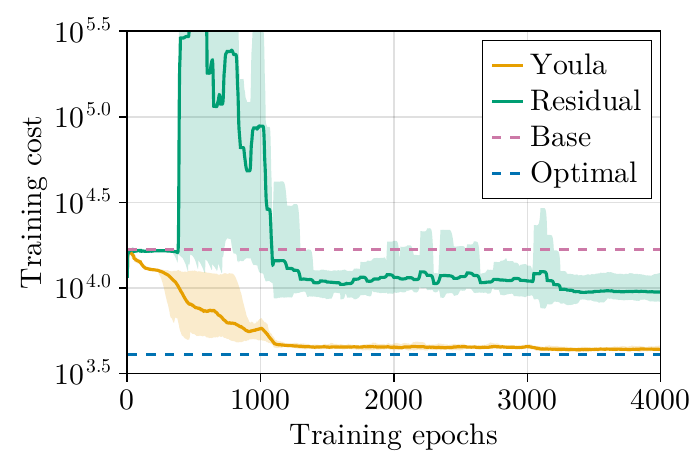}
        \caption{Mean training cost based on short-horizon rollouts.}
        \label{fig:doyle-example-traincurve}
    \end{subfigure}
    \begin{subfigure}[b]{\linewidth}
        \centering
        \includegraphics[trim={0.2cm 0.4cm 0cm 8cm},clip,width=0.75\linewidth]{Images/stability_guarantees_learning_singlecol.pdf}
        \caption{Mean test cost for long-term policy rollouts.}
        \label{fig:doyle-example-rollouts}
    \end{subfigure}
    \caption{Youla-REN achieves long-term stability even with short-horizon training on the classic LQG problem from \cite{Doyle1978}. Bands in (a) show the range across 10 random seeds. Lines in (b) show test cost rollouts over time of particular policies.}
    \label{fig:doyle-example}
\end{figure}

For comparison, we also trained \emph{Residual-REN} policies
\begin{equation}\label{eqn:residual-policy}
    u=\mathcal{K}_b(y)+\mathcal{Q}(y),
\end{equation}
where $\mathcal{K}_b$ is the stabilizing base controller and the learnable component $\mathcal{Q}$ is parameterized by a REN. As shown in Fig.~\ref{fig:rl-architectures}, the main difference from Youla-REN is that the system $\mathcal{Q}$ in a Residual-REN acts directly on measurements $y$ rather than the innovations $\tilde{y}$. A direct consequence is that Youla-REN guarantees closed-loop stability (provided that $\mathcal{Q}$ is contracting and Lipschitz) while Residual-REN does not.

The training curves in Fig.~\ref{fig:doyle-example-traincurve} show that the Youla-REN policies consistently outperform the Residual-RENs and smoothly converge to a near-optimal final cost. In contrast, the Residual-RENs never approach the optimal LQG cost and yield large variance during training. We found that the Residual-RENs were highly sensitive to the learning rate, with slightly smaller values resulting in almost no improvement over the base controller, while larger values lead to even greater fluctuations and a worse final cost. Since the true optimal policy is near the boundary of the stabilizing controller set, it is common to encounter unstable Residual-REN policies during training (e.g., at epoch 687 in the top panel of Fig.~\ref{fig:doyle-example-rollouts}). This is not possible with the Youla-REN due to its built-in stability guarantees.

We also compare the long-term test performance of each policy architecture in Fig.~\ref{fig:doyle-example-rollouts}. Examining Residual-RENs at training epochs 1500 and 4000, the policies appear to stabilize the closed-loop system \textit{over the training horizon} (i.e., the cost is less than the base cost). However, rolling out these policies over longer test horizons reveals unstable responses. The Youla-REN demonstrates consistently strong performance over long time periods at all stages during and after training.

\subsection{Robust Stability with Model Uncertainty} \label{sec:exp-uncertainty}

Our final example demonstrates that Youla-REN policies can achieve good performance while maintaining closed-loop stability in the presence of model parameter uncertainty. Consider the continuous-time, linearized cart-pole system from \cite{Wang++2022} with zero-mean Gaussian process and measurement noise variances $\sigma_w = 10^{-3}$, $\sigma_v = 10^{-4}$, respectively. We discretized the system with the forward-Euler method ($\Delta t = 0.02$\,s) and chose a pole length of 10\,m, cart mass of 1\,kg, and pole mass $m_p \in [0.14, 0.35]$\,kg. The control objective was to minimize an infinite-horizon quadratic cost with weights $Q = \texttt{diag}(1, 0, 1, 0)$ and $R = I$. We chose a discrete-time, time-invariant LQG controller as the base controller, designed at a nominal mass $m_p^\star=0.2$. We computed its control gain with the same $Q,R$ as the cost function, but the steady-state Kalman filter was designed with covariances $\Sigma_w = \texttt{diag}(1, 10^3, 1, 10^3) \times \Delta t$ and $\Sigma_v = 10^{-3} I / \Delta t$ for the process and measurement noises, respectively, to ensure closed-loop stability across the whole uncertain $m_p$ range. Neural policies were trained on batches of 64 randomly sampled $m_p$ within the given range and random initial states.

\begin{figure}[!t]
    \centering
    \hspace{-2mm}
    \begin{subfigure}[b]{0.46\linewidth}
        \centering
        \includegraphics[width=\linewidth]{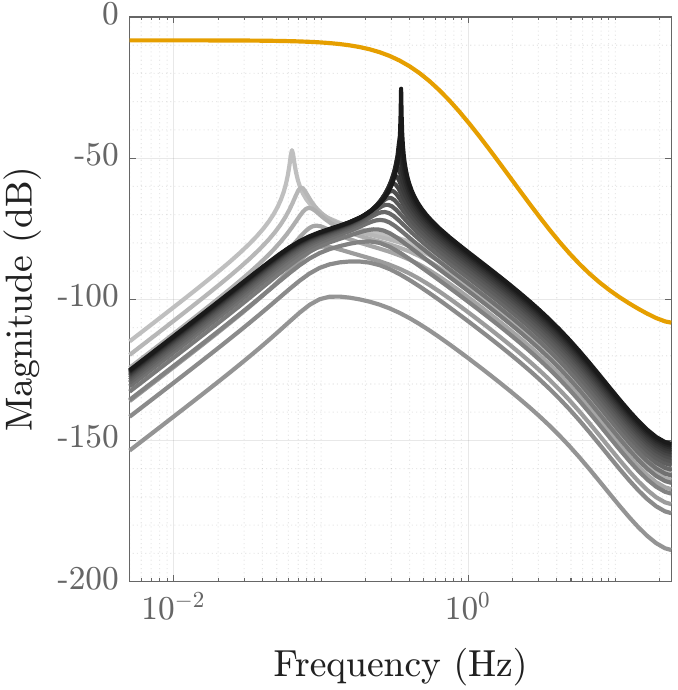}
        \caption{$\tilde{u} \mapsto \tilde{y}$}
        \label{fig:lcp-filter-udy}
    \end{subfigure}
    \hspace{5mm}
    \begin{subfigure}[b]{0.46\linewidth}
        \centering
        \includegraphics[width=\linewidth]{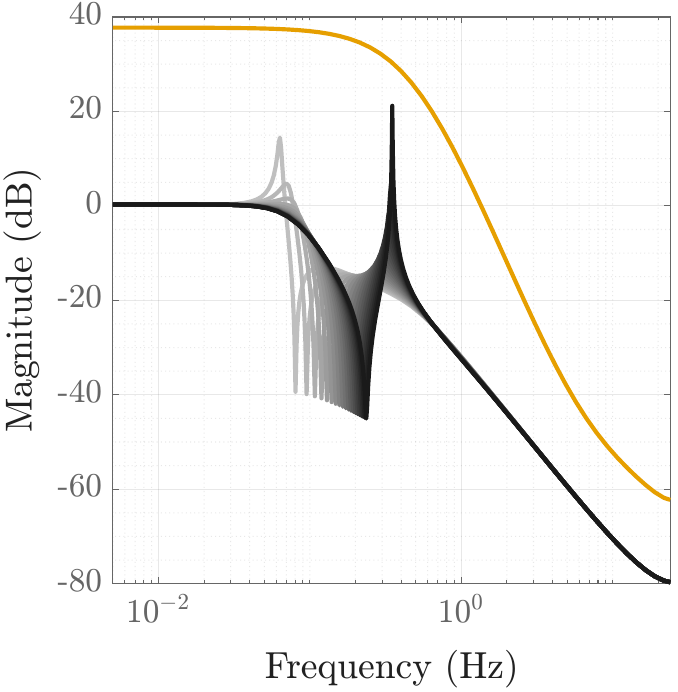}
        \caption{$\tilde{u} \mapsto y$}
        \label{fig:lcp-filter-uy}
    \end{subfigure}
    \caption{Bode magnitude plots for discrete-time, linear cart-pole under the base LQG controller with different pole masses $m_p$, where the color gradient darkens as $m_p$ increases. Yellow lines show the inverse of the (discretized) weighting filters \eqref{eqn:weighting-filter}.}
    \label{fig:lcp-weighting-filters}
\end{figure}

\begin{figure*}[!t]
    \centering
    \begin{subfigure}[b]{0.27\textwidth}
        \centering
        \includegraphics[trim={0cm 0cm 23cm 0cm},clip,height=10\baselineskip]{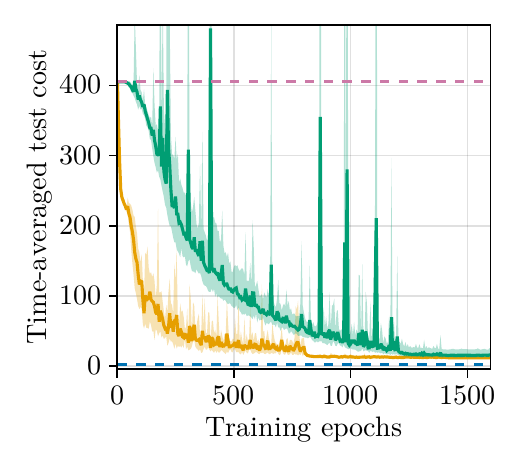}
        \caption{Training curves.}
        \label{fig:lcp-costs-traincurves}
    \end{subfigure}
    \hfill
    \begin{subfigure}[b]{0.27\textwidth}
        \centering
        \includegraphics[trim={8.6cm 0cm 14.8cm 0cm},clip,height=10\baselineskip]{Images/lcp_youla_residual_costs_fullwidth.pdf}
        \caption{Test cost rollouts.}
        \label{fig:lcp-costs-rollouts}
    \end{subfigure}
    \hfill
    \begin{subfigure}[b]{0.44\textwidth}
        \centering
        \includegraphics[trim={17cm 0cm 0cm 0cm},clip,height=10\baselineskip]{Images/lcp_youla_residual_costs_fullwidth.pdf}
        \caption{Sensitivity to model uncertainty.}
        \label{fig:lcp-costs-sensitivity}
    \end{subfigure}
    \caption{Mean test costs for the cart-pole problem with uncertain pole mass: (a) short-horizon test cost during training (evaluated at 2 $\times$ training horizon); (b) test cost for long-term rollouts of particular policies; and (c) long-horizon test cost as a function of the uncertain pole mass (evaluated at 8 $\times$ training horizon). LQG refers to the optimal time-varying implementation; LTI LQG refers to a steady-state Kalman filter combined with an LQR gain. Black-box LSTM policies are discussed in Section~\ref{sec:black-box}.}
    \label{fig:lcp-costs}
\end{figure*}

We parameterized robustly-stabilizing policies for this system with Youla-$\gamma$REN, a Lipschitz-bounded variant of Youla-REN. We chose the Lipschitz bound $\gamma$ using a small-gain approach with loop shaping to enable both closed-loop stability and strong performance (Prop.~\ref{prop:iqc}). For frequency weightings, we chose the output filter to be $\mathcal{W}_1(s)=1$ and the input filter
\begin{equation} \label{eqn:weighting-filter}
    \mathcal{W}_2(s) = \frac{(s + 3)^4}{\nu (s + 50)^4}
\end{equation}
with $\nu$ given in Tab.~\ref{tab:bounds-and-gains}. The inverse of $\mathcal{W}_2(s)$ (Fig.~\ref{fig:lcp-filter-udy}) serves as a good heuristic to upper bound the base-controlled system with uncertain pole mass. We then chose $\gamma$ as the inverse of the $\mathcal{H}_\infty$ norm of the base-controlled system $\tilde{u} \mapsto \tilde{y}$ with filtering. 
Based on an empirical study, we observed only minor differences in the results when swapping $\mathcal{W}_1$ and $\mathcal{W}_2$.
Note that the shape of $\mathcal{W}_2(s)$ allows for higher gains at moderately high frequencies, which can help to learn a policy that rapidly corrects for displacements of the pole angle. We did not shape $\mathcal{W}_2$ at low frequencies, since low-frequency poles require training over very long horizons to capture their behavior.

For comparison, we also trained Residual-LSTM policies -- the residual RL architecture \eqref{eqn:residual-policy} with $\mathcal{Q}$ parameterized by a Long Short-Term Memory network (LSTM) \cite{Hochreiter+Schmidhuber1997}. LSTMs are popular recurrent networks known to perform well on RL tasks \cite{Bakker2001}. We found that Residual-LSTMs were easier to train for this problem than unconstrained Residual-RENs.

\subsubsection{Results \& discussion} 

\begin{table}[!b]
    \centering
    \caption{Policy hyperparameters.}
    \begin{tabular}{l|c|c} 
    \toprule
    \textbf{Policy Architecture} &  Param. $\nu$ in $\mathcal{W}_2$ & Lip. bound $\gamma$ of $\mathcal{Q}$ \\
    \midrule
    Youla-$\gamma$REN & $5\times 10^{-4}$ & 1.7 \\
    Youla-$\gamma$REN (linear) & $5\times 10^{-4}$ & 1.7 \\
    Youla-$\gamma$REN (no filter)   &     - & 120 \\
    Residual-$\gamma$REN  &    $10^{-2}$ & 0.15 \\
    Residual-LSTM                        & - & - \\
    \bottomrule
    \end{tabular}
    \label{tab:bounds-and-gains}
\end{table}

We first compare the Youla-$\gamma$REN and Residual-LSTM in Figs.~\ref{fig:lcp-costs-traincurves} and \ref{fig:lcp-costs-rollouts} to investigate the effect of stability guarantees on policy training and long-term test rollouts with model uncertainty. We observe similar qualitative results to Fig.~\ref{fig:doyle-example}. Specifically, the Youla-$\gamma$RENs quickly and smoothly converge to a near-optimal cost during training, and yield stable closed-loop responses over much longer test horizons than the training horizon. In contrast, the Residual-LSTMs often converge to unstable policies during training, causing spikes in their cost curves, and also exhibit closed-loop instability at test time when rolled out over long horizons (e.g., 6 times the training horizon in Fig.~\ref{fig:lcp-costs-rollouts}).

The closed-loop responses under different pole masses (the nominal $m_p^\star$ and two end points) are depicted in Fig.~\ref{fig:lcp-uncertainty-trajectories}. Both the Youla-$\gamma$REN and Residual-LSTM policies can stabilize the system with a nominal and large pole mass. However, when the pole mass is small, the Residual-LSTM exhibits marginally unstable behavior. That is, the policy rollout of Residual-LSTM is close to that of Youla-$\gamma$REN within the training horizon, but quickly diverges as the time increases. This highlights the importance of model-based stability guarantees in deep RL to achieve reliable long-term test performance.

The sensitivity of the learned policies to variations in the uncertain parameter is shown further in Fig.~\ref{fig:lcp-costs-sensitivity}. The Residual-LSTM performs well over most of the range, but exhibits instability at small pole masses. The proposed Youla-$\gamma$REN ensures stability and achieves near optimal performance across the whole parameter range. Although its cost is greater than the (optimal) time-varying LQG with known mass, it is very close the time-invariant LQG with known mass, which is reasonable as our policy is also time-invariant. This performance cannot simply be achieved by an LQG controller designed at the nominal mass (shown in gray in Fig.~\ref{fig:lcp-costs-sensitivity}) or a Youla-$\gamma$REN controller with linear $\mathcal{Q}$ (see Fig.~\ref{fig:lcp-ablation}). This suggests that the proposed nonlinear Youla-REN controller has the ability to learn to adapt to different model parameters.

\begin{table}[!b]
\centering
\caption{Policy architectures for the ablation study (Fig.~\ref{fig:lcp-ablation}).}
\begin{tabular}{l|ccc} 
\toprule
Policy &  Architecture & $\mathcal{Q}$ & Weight filter \\
\midrule
Youla-$\gamma$REN (linear) & Youla & Linear & Yes \\
Youla-$\gamma$REN (no filter)   & Youla & Nonlinear & No \\
Residual-$\gamma$REN  &  Residual & Nonlinear & Yes \\
\bottomrule
\end{tabular}
\label{tab:policy}
\end{table}

\begin{figure}[!t]
    \centering
    \hspace{-5mm}
    \includegraphics[width=\linewidth]{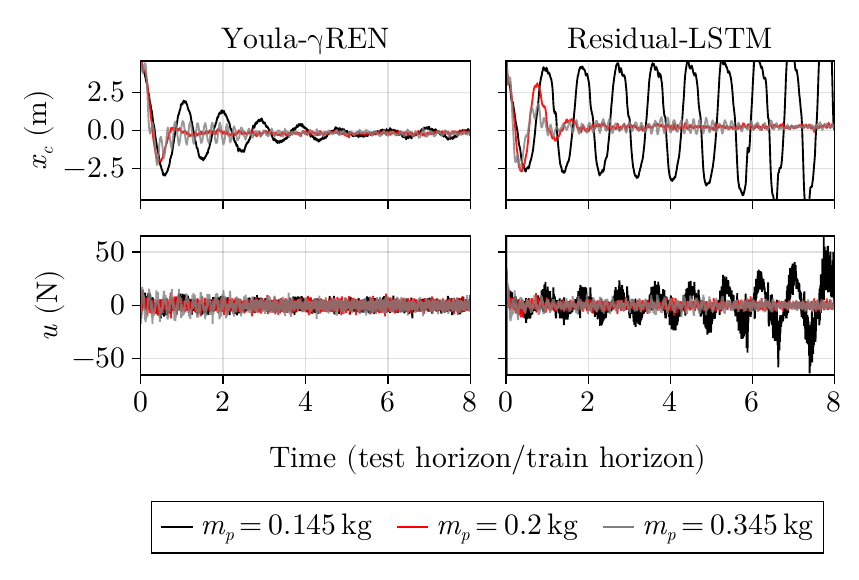}
    \caption{Policy rollouts on linear cart pole with the same initial state and different pole mass $m_p$, where $x_c$ is the cart position. }
    \label{fig:lcp-uncertainty-trajectories}
\end{figure}

\subsubsection{Ablation study} 

We now conduct an ablation study on the Youla-$\gamma$REN policy class. We trained policies with the three architectures in Tab.~\ref{tab:policy}, all of which have built-in stability guarantees via proper choice of the hyperparameters in Tab~\ref{tab:bounds-and-gains}.

Fig.~\ref{fig:lcp-ablation} shows that each of the three Youla-$\gamma$REN components (the Youla parameterization, nonlinear $\mathcal{Q}$, and frequency-weighted filter) is crucial to learn high-performing policies. The linear Youla-$\gamma$REN plateaus at a much higher cost than its nonlinear counterparts due to the poor expressivity of a linear $\mathcal{Q}$. When the weighting filter is removed, the gain of the Youla-$\gamma$REN is restricted equally across all frequencies, leading to a higher cost ceiling. Perhaps most interestingly, there is an extreme drop in performance when switching from the Youla configuration to Residual RL. The main reason is that the gain of $\tilde{u} \mapsto \tilde{y}$ is much smaller than $\tilde{u} \mapsto y$ (Fig.~\ref{fig:lcp-weighting-filters}), allowing for a more expressive (i.e. larger Lipschitz bound) nonlinear $\mathcal{Q}$ when ensuring closed-loop stability via the small-gain theorem.
This reinforces the fact that the policy architecture does matter in deep RL \cite{Roberts++2011}. We learn performant policies with built-in robustness to model uncertainty by combining all three key elements of the Youla-$\gamma$REN policy architecture.

\begin{figure}[!t]
    \centering
    \hspace{-5mm}
    \includegraphics[width=0.95\linewidth]{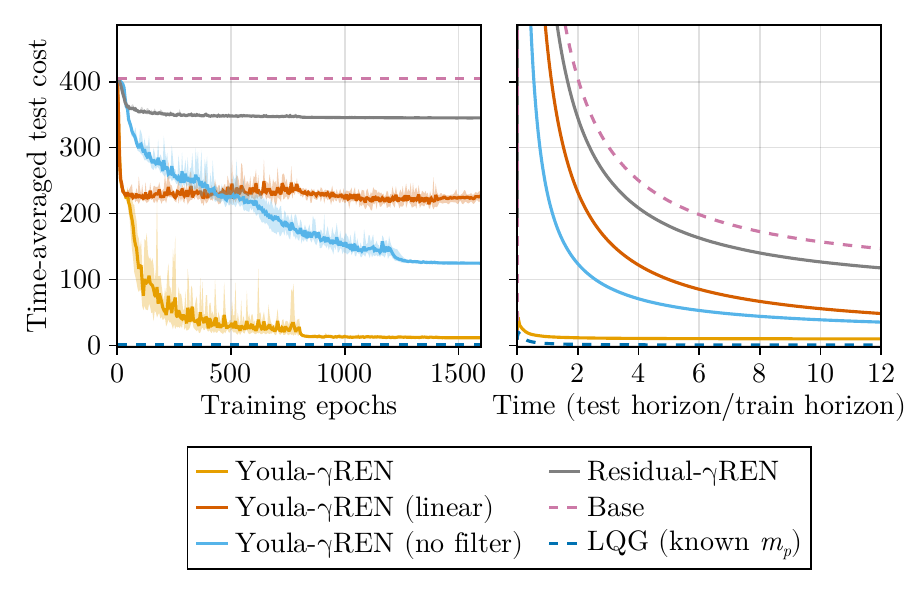}
    \caption{Ablation study on the Youla-$\gamma$REN. }
    \label{fig:lcp-ablation}
\end{figure}

\subsubsection{Black-box Policies} 
\label{sec:black-box}

We also compared the Youla-REN with Multi-Layer Perceptron (MLP) and LSTM policies in the black-box RL architecture from Fig.\ref{fig:feedback-architecture}. The training curves in Fig.~\ref{fig:lcp-blackbox} clearly demonstrate the following: black-box policies do not reliably converge on stabilizing controllers for this problem; in cases where LSTM policies converge below the base cost, they achieve a similar final cost to the Youla-REN only after many more training epochs; both MLP and LSTM policies exhibit large spikes throughout training, even after convergence, as they frequently stumble upon de-stabilizing policies; and black-box LSTM policies exhibit instability at test time when rolled out beyond the training horizon, just like the Residual-LSTMs (see Fig.~\ref{fig:lcp-costs-rollouts}).

Note that the results in Figs.~\ref{fig:lcp-blackbox} show the best-performing policies after performing a sweep over a range of learning rates. It is possible that with longer training horizons, more random batches, different optimization algorithms, and further hyperparameter tuning, one could train stabilizing black-box policies for this task. The advantage of our approach is that the training process is completely decoupled from the policy's stability certificate -- that is, closed-loop stability is guaranteed by construction, regardless of the training setup.

\begin{figure}[!t]
    \centering
    \hspace{-5mm}
    \includegraphics[width=0.95\linewidth]{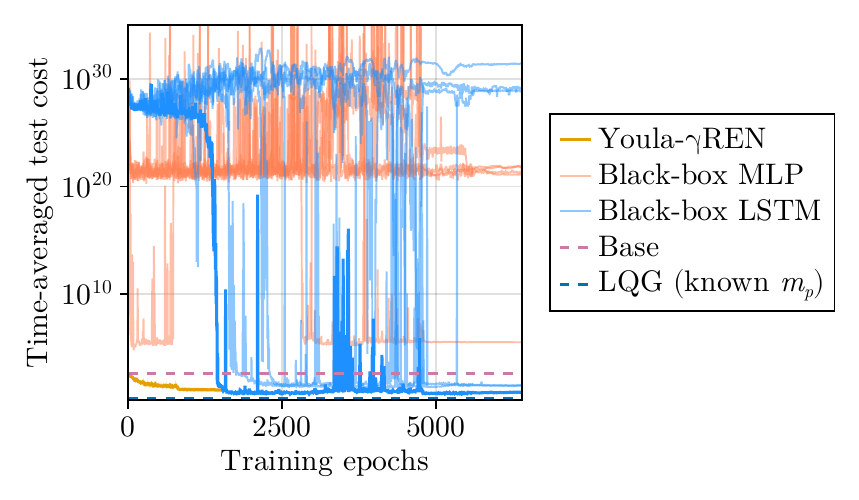}
    \caption{Comparison of Youla-$\gamma$REN to black-box policies over many random model initializations. Test rollouts for the best-performing LSTM (highlighted) are shown in Fig.~\ref{fig:lcp-costs-rollouts}.}
    \label{fig:lcp-blackbox}
\end{figure}

%% file: proofs.tex
%
%

\subsection{Useful Lemma}

Consider a time-varying nonlinear system
\begin{equation}\label{eqn:additive-perturbation}
    x_{t+1}=f(x_t,t)+ d_t
\end{equation}
with $x_t,d_t\in \R^n$. Note that $d_t$ could also be state-dependent. Let $x^\star$ be a disturbance-free trajectory with $d_t= 0\ \forall t\in\mathbb{N}$. 
\begin{lemma} \label{lem:additive-d}
    Suppose $x_{t+1} = f(x_t,t)$ is contracting with rate $\alpha$ and overshoot $\beta$. 
    If the disturbance satisfies $\abs{d_t} \le \bar{d}+ \delta \varepsilon^t $ with $\bar{d},\delta\in\mathbb{R}^+$ and $ \varepsilon\in[0,1)$, then 
    \begin{equation} \label{eqn:exp-conv-dist1}
        \abs{x_t - x^\star_t} \le \beta \rho^t (\abs{x_0 - x^\star_0} + \delta \tau)+\frac{\beta}{1-\alpha}\bar{d}, \quad \forall t\in \mathbb{N}
    \end{equation}
    where $\rho \in [0,1)$ and $\tau \in \mathbb{R}^+$ depend on $\alpha, \varepsilon$.
\end{lemma}
\begin{proof}
    From \cite[Lemma A.1]{barbara2026thesis} we can obtain 
    \begin{equation}
        \begin{split}
            |\Delta x_t|&\leq \beta \alpha^t|\Delta x_0|+\beta \sum_{k=0}^{t-1}\alpha^{t-1-k}|d_k| \\
            &\leq \beta \alpha^t|\Delta x_0| +\frac{\beta}{1-\alpha}\bar{d}+\beta\delta\sum_{k=0}^{t-1}\alpha^{t-1-k}\varepsilon^k
        \end{split}
    \end{equation}
    with $\Delta x=x-x^\star$. When $\alpha\neq \varepsilon$, we have
    \[
    \sum_{k=0}^{t-1}\alpha^{t-1-k}\varepsilon^k=\frac{\alpha^t-\varepsilon^t}{\alpha-\varepsilon}\leq \tau \rho^t
    \]
    with $\rho=\max(\alpha,\varepsilon)$ and $\tau=1/|\alpha-\varepsilon|$. When $\alpha=\epsilon$, we have 
    \[
    \sum_{k=0}^{t-1}\alpha^{t-1-k}\varepsilon^k=t\alpha^{t-1}\leq \tau \rho^t
    \]
    with $\rho\in (\alpha, 1)$ and $\tau=\max_{t\in \mathbb{N}}\,t\alpha^{t-1}/\rho^t$.
\end{proof}

\subsection{Proof of Theorem~\ref{thm:youla-disturbance-tube}} \label{app:proof-main-result}
 
First, we show that the bound of the innovations signal $\tilde{y}$ \eqref{eqn:innovations} is independent of $\tilde{u}$, despite there being a feedback loop between $\tilde{u}$ and $\tilde{y}$ via $\mathcal{Q}$. Applying observer correctness (\ref{assump:observer-correctness}) to the nonlinear system $\mathcal{G}$, we have
\begin{equation}\label{eqn:sys-obs}
    \begin{split}
        x_{t+1}&=f(x_t,\eta_t, u_t)+w_t \\
        &=f_o(x_t,\eta_t,u_t, h(x_t))+w_t \\
        &=f_o(x_t,\eta_t, u_t, y_t-v_t)+w_t \\
        &=f_o(x_t,\eta_t, u_t, y_t) + \tilde{d}_t
    \end{split}
\end{equation}
with $\tilde{d}_t:=f_o(x_t,\eta_t, u_t, y_t-v_t)-f_o(x_t,\eta_t, u_t, y_t)+w_t$. From Assumption (\ref{assump:observer-contract}) we have that the observer
\begin{equation}
    \hat{x}_{t+1}=f_o(\hat{x}_t,\eta_t,u_t,y_t)
\end{equation}
is contracting with rate $\alpha_o$ and overshoot $\beta_o$ for some $\alpha_o\in [0,1)$ and $\beta_o\in \R^+$. Moreover, $(\eta,u,y)\mapsto \hat{x}$ is $\gamma_o$-Lipschitz for some $\gamma_o\in\R^+$. Then, we can have
\begin{equation}
    |\tilde{d}_t|\leq \gamma_o|v_t|+|w_t| \leq \gamma_o'|d_t|
\end{equation}
with $d_t=[w_t; v_t]$ and $\gamma_o'=\sqrt{\gamma_o^2+1}$. From Lemma~\ref{lem:additive-d} and \cite[Lemma A.2]{barbara2026thesis}, the observer error $\tilde{x}=x-\hat{x}$ satisfies
\begin{align}
        |\tilde{x}_t| &\le  \beta_o \alpha_o^t |\tilde{x}_0|+\sigma_o\gamma_o'\|d\|_{t-1,\infty} \label{eqn:obs-x-t}\\
    \norm{\tilde{x}}_T &\le \sigma_o\gamma_o' \norm{d}_T + \sigma_o'|\tilde{x}_0| \label{eqn:obs-x-norm}
\end{align} 
where $\sigma_o=\beta_o/(1-\alpha_o) $ and $\sigma_o'=\beta_o/\sqrt{1-\alpha_o^2}$. Moreover, since the output map $h$ is Lipschitz, the innovation signal $\tilde{y}:=y-h(\hat{x})=h(x)-h(\hat{x})+v$ satisfies
\begin{align}
        |\tilde{y}_t| &\le  \gamma_h\beta_o \alpha_o^t |\tilde{x}_0|+(\gamma_h\sigma_o\gamma_o'+1)\|d\|_{t,\infty} \label{eqn:obs-y-t}\\
    \norm{\tilde{y}}_T &\le \left(\gamma_h\sigma_o\gamma_o'+1\right) \norm{d}_T + \gamma_h\sigma_o'|\tilde{x}_0|\label{eqn:obs-y-norm}
\end{align}
where $\gamma_h\in \R^+$ is a Lipschitz bound of $h$ w.r.t. $x$. 

\subsubsection{d-tube contraction} We consider two closed-loop state trajectories $\chi^a, \chi^b$ with  under the same input $(\eta, d)$, where $\chi_t^{i}=[\bar{x}_t^{i}; \tilde{x}_t^{i}]$ for $i\in \{a,b\}$. From \eqref{eqn:obs-x-t} we have 
\begin{equation}\label{eqn:delta-x-tilde}
    |\Delta \tilde{x}_t|\leq |\tilde{x}_t^a|+|\tilde{x}_t^b|\leq \beta_o\alpha_o^t(|\tilde{x}_0^a|+|\tilde{x}_0^b|) + 2\sigma_o\gamma_o'\|d\|_{t-1,\infty}
\end{equation}
Similarly, from \eqref{eqn:obs-y-t} we can obtain 
\begin{equation}
    |\Delta \tilde{y}_t|\leq |\tilde{y}_t^a|+|\tilde{y}_t^b|\leq
    \tilde{\beta}_o\alpha_o^t(|\tilde{x}_0^a|+|\tilde{x}_0^b|)+\tilde{\sigma}_o\|d\|_{t,\infty}
\end{equation}
with $\tilde{\beta}_o=\gamma_h\beta_o$ and $\tilde{\sigma}_o=2(\gamma_h\sigma_o\gamma_o'+1)$. We then consider the system $ \mathcal{G}_{\mathcal{K}_b}\circ \mathcal{Q}  $, which is contracting and Lipschitz by assumptions. By denoting $\bar{x}$ as the state of $ \mathcal{G}_{\mathcal{K}_b}\circ \mathcal{Q}  $, we have
\begin{equation}
    \begin{split}
    \bar{x}_{t+1}^a&=f_c(\bar{x}_{t}^a,\eta_t,d_t,\tilde{y}_t^a)
        =f_c(\bar{x}_{t}^a,\eta_t,d_t,\tilde{y}_t^b)+\Delta f_{c,t}, \\
    \bar{x}_{t+1}^b&=f_c(\bar{x}_{t}^b,\eta_t,d_t,\tilde{y}_t^b).
    \end{split}
\end{equation}
Since $f,f_b,f_q$ are Lipschitz w.r.t. their system inputs, we have that $f_c$ is $\tilde{\gamma}_c$-Lipschitz w.r.t. $\tilde{y}$ for some $\tilde{\gamma}_c\in \R^+$, leading to $|\Delta f_{c,t}|\leq \tilde{\gamma}_{c} |\Delta \tilde{y}_t|$.  From Lemma~\ref{lem:additive-d}, we have
\begin{equation}\label{eqn:delta-x-bar}
    \begin{split}
    |\Delta \bar{x}_t|&\leq \beta_c\rho_c^t(|\Delta \bar{x}_0|+\tilde{\gamma}_c\tilde{\beta}_o \tau_c(|\tilde{x}_0^a|+|\tilde{x}_0^b|))\\
    &\;+ \tilde{\gamma}_c\tilde{\sigma}_o\sigma_c \|d\|_{t-1,\infty}
    \end{split}
\end{equation}
with $\sigma_c=\beta_c/(1-\alpha_c)$, $\rho_c\in [0,1)$ and $ \tau_c\in \R^+$ depend on $\alpha_c, \alpha_o$, where $\alpha_c,\beta_c$ are the contracting rate and overshoot of $\mathcal{G}_{\mathcal{K}_b}\circ \mathcal{Q}$, respectively. Combining \eqref{eqn:delta-x-tilde} and \eqref{eqn:delta-x-bar} yields \eqref{eqn:tube-contraction} with $\alpha=\rho_c$, $\gamma_1=2\sigma_o\gamma_o'+\tilde{\gamma}_c\tilde{\sigma}_o\sigma_c$ and
\begin{equation}
    \kappa_1(\chi_0^a,\chi_0^b)=\bar{\gamma}_1\abs{ \bar{x}_0^a-\bar{x}_0^b} + \tilde{\gamma}_1(|\tilde{x}_0^a| + |\tilde{x}_0^b|)
\end{equation}
with $\bar{\gamma}_1=\beta_c$ and $\tilde{\gamma}_1=\beta_o+\tilde{\gamma}_{c}\beta_c\tilde{\beta}_o\tau_c$.

\subsubsection{d-tube Lipschitzness} We consider two closed-loop trajectories $(\chi^a,\eta^a,d^a,z^a)$, $(\chi^b,\eta^b,d^b,z^b)$. From \eqref{eqn:obs-y-norm} we have
\begin{equation}\label{eqn:delta-yt-norm}
    \begin{split}
        \|\Delta\tilde{y}\|_T&\leq \|\tilde{y}^a\|_T+\|\tilde{y}^b\|_T\\
        &\leq \tilde{\gamma}_o(\|d^a\|_T+\|d^b\|_T)+\bar{\gamma}_o(|\tilde{x}_0^a|+|\tilde{x}_0^b|)
    \end{split}
\end{equation}
with $\tilde{\gamma}_o=\gamma_h\sigma_o\gamma_o'+1$ and $\bar{\gamma}_o=\gamma_h\sigma_o'$. Since $\mathcal{G}_{\mathcal{K}_b}\circ \mathcal{Q}:(\eta,d,\tilde{y})\mapsto z $ is contracting and Lipschitz, we can show that
\begin{equation}\label{eqn:delta-z-norm}
    \|\Delta z\|_T\leq \gamma_c(\|\Delta \eta\|_T+\|\Delta d\|_T+\|\Delta \tilde{y}\|_T)+\mu_c |\Delta \chi_0| 
\end{equation}
with $\mu_c=\gamma_{h_c}\beta_c/\sqrt{1-\alpha_c^2}$, where $\gamma_c, \gamma_{h_c}$ are the Lipschitz bounds of system $\mathcal{G}_{\mathcal{K}_b}\circ \mathcal{Q}$ and its output map $h_c$. Substituting \eqref{eqn:delta-yt-norm} into \eqref{eqn:delta-z-norm} yields \eqref{eqn:tube-lipschitz} with $\gamma=\gamma_c$, $\gamma_2=\gamma_c\tilde{\gamma}_o$ and
\begin{equation}
    \kappa_2(\chi_0^a,\chi_0^b)=\bar{\gamma}_2\abs{ \bar{x}_0^a-\bar{x}_0^b} + \tilde{\gamma}_2(|\tilde{x}_0^a| + |\tilde{x}_0^b|)
\end{equation}
with $\bar{\gamma}_2=\mu_c$ and $\tilde{\gamma}_2=\gamma_c\bar{\gamma}_o+\mu_c$.

\subsection{Proof of Proposition~\ref{prop:innov-system}}
First consider the LTI system \eqref{eqn:linear-system} with the linear observer from \eqref{eqn:linear-Kb}. The innovations depend only on the observer error $\tilde{x}$ and external disturbances, since
\begin{equation*}
    \begin{aligned}
        \tilde{x}_{t+1} &= (A - LC) \tilde{x}_t - L v_t + w_t \\
        \tilde{y}_t &= C \tilde{x}_t + v_t.
    \end{aligned}
\end{equation*}
This system is contracting and Lipschitz and of the form \eqref{eqn:innovations-system}.

Next, consider the fully-observed nonlinear system \eqref{eqn:system-state-feedback} with innovations \eqref{eqn:state-feedback-innovations}. We can re-write \eqref{eqn:state-feedback-innovations} as 
\begin{equation*}
    \psi_{t+1} = w_t, \quad \tilde{y}_t = \psi_t
\end{equation*}
which is also contracting and Lipschitz and of the form \eqref{eqn:innovations-system}.

Finally, consider the partially-observed nonlinear system \eqref{eqn:system} with the observer \eqref{eqn:observer}. Under zero disturbances ($d \equiv 0$, $\tilde{x}_0 \equiv 0$), by observer correctness (\ref{assump:observer-correctness}) the observer is an exact replica of the plant, hence  $\hat{x}_t=x_t \ \forall\, t\in \mathbb{N}$. Therefore $\tilde{y}_t = 0\ \forall\, t\in \mathbb{N}$, which is a trivial case of \eqref{eqn:innovations-system}.

\subsection{Proof of Theorem~\ref{thm:youla-decoupled}} \label{app:proof-decoupled}
Since all of $\mathcal{T}$, $\mathcal{Q}$, and $\mathcal{G}_{\mathcal{K}_b}$ are contracting and Lipschitz by assumption, then the result follows by the series composition properties of contracting systems and Lipschitz mappings.

\subsection{Proof of Proposition~\ref{prop:iqc}} \label{app:proof-iqc}

The proof is provided in \cite[Chapter~5]{barbara2026thesis} and is a straightforward application of $(Q, S, R)$-dissipativity theory for networked systems in the incremental setting.

\subsection{Proof of Theorem~\ref{thm:partial-converse} \& Associated Corollaries} \label{app:proof-partial-converse}

The closed-loop system \eqref{eqn:innov-obsv-base-ctrl} maps $(\eta,\tilde{y}) \mapsto \hat{z}$ according to
\begin{equation} \label{eqn:cl-obsv-base-K}
    \begin{aligned}
        \hat{x}_{t+1} &= f_o(\hat{x}_t, \eta_t, h_\mathcal{K}(\phi_t, \eta_t, h(\hat{x}_t) + \tilde{y}_t), h(\hat{x}_t) + \tilde{y}_t) \\
        \phi_{t+1} &= f_\mathcal{K}(\phi_t, \eta_t, h(\hat{x}_t) + \tilde{y}_t) \\
        \hat{z}_t &= [\hat{x}_t; h_\mathcal{K}(\phi_t, \eta_t, h(\hat{x}_t) + \tilde{y}_t)],
    \end{aligned}
\end{equation}
and will be used in the proofs of both corollaries.

\subsubsection{Theorem~\ref{thm:partial-converse}}

We show that $\mathcal{K}$ can be re-written in the form \eqref{eqn:youla-ctrl} with a contracting and Lipschitz Youla parameter $\mathcal{Q}_\mathcal{K}$. Augment $\mathcal{K}$ with a contracting and Lipschitz base controller \eqref{eqn:base-controller} and the observer \eqref{eqn:innov-obsv-base-ctrl-observer}. Using  $y_t = \tilde{y}_t + h(\hat{x}_t)$ and $\tilde{u}_t = u_t - \hat{u}_t$, the resulting system $\mathcal{Q}_\mathcal{K}: (\eta, \tilde{y}) \mapsto \tilde{u}$ has dynamics 
\begin{equation}
    \begin{aligned}
        \hat{x}_{t+1} &= f_o(\hat{x}_t, \eta_t, h_\mathcal{K}(\phi_t, \eta_t, h(\hat{x}_t) + \tilde{y}_t), h(\hat{x}_t) + \tilde{y}_t) \\
        s_{t+1} &= f_b(s_t, \eta_t, h_\mathcal{K}(\phi_t, \eta_t, h(\hat{x}_t) + \tilde{y}_t), h(\hat{x}_t) + \tilde{y}_t) \\
        \phi_{t+1} &= f_\mathcal{K}(\phi_t, \eta_t, h(\hat{x}_t) + \tilde{y}_t) \\
        \tilde{u}_t &= h_\mathcal{K}(\phi_t, \eta_t, h(\hat{x}_t) + \tilde{y}_t) - k(s_t, \eta_t, h(\hat{x}_t) + \tilde{y}_t).
    \end{aligned}
\end{equation}
This system $\mathcal{Q}_\mathcal{K}$ is a series interconnection of $\mathcal{K}_b$ and \eqref{eqn:innov-obsv-base-ctrl}, both of which are contracting and Lipschitz by assumption (see Fig.~\ref{fig:youla-parameter}). Noting that the static output map $h$ is Lipschitz by assumption~\ref{assump:smooth-functions}, then $\mathcal{Q}_\mathcal{K}$ must also be contracting and Lipschitz by the series composition properties of contracting systems and Lipschitz mappings.

Moreover, if $\mathcal{Q}=\mathcal{Q}_\mathcal{K}$ in the Youla architecture \eqref{eqn:youla-ctrl}, then the control signal $u_t = \hat{u}_t + h_\mathcal{K}(\phi_t, \eta_t, y_t) - \hat{u}_t$ is unchanged from $u = \mathcal{K}(\eta, y)$. Hence $\mathcal{K}$ can be written in the form \eqref{eqn:youla-ctrl} with $\mathcal{Q}=\mathcal{Q}_\mathcal{K}$ contracting and Lipschitz.

\subsubsection{Corollary~\ref{cor:partial-conv-special-cases}}
The closed-loop system $(\eta, w) \mapsto z$ consisting of \eqref{eqn:system-state-feedback} in feedback with $u = \mathcal{K}(\eta, y)$ \eqref{eqn:generic-ctrl} has dynamics
\begin{equation} \label{eqn:cl-perturbed-state-feedback}
\mathcal{K}:\; \left\{
\begin{aligned}
    x_{t+1} &= f(x_t, \eta_t, h_\mathcal{K}(\phi_t, \eta_t, x_t)) + w_t \\
    \phi_{t+1} &= f_\mathcal{K}(\phi_t, \eta_t, x_t) \\
    z_t &= [x_t; h_\mathcal{K}(\phi_t, \eta_t, x_t)].
\end{aligned}\right.
\end{equation}
This system is contracting and Lipschitz by assumption. Make the substitutions $x_t = \hat{x}_t + w_{t-1}$ and $\tilde{y}_t = w_{t-1}$ which follow by \eqref{eqn:innovations}, \eqref{eqn:state-feedback-innovations} since $y_t = x_t$. Then \eqref{eqn:cl-perturbed-state-feedback} becomes
\begin{equation} \label{eqn:cl-perturbed-state-feedback-transformed}
\begin{aligned}
    \hat{x}_{t+1} &= f(\hat{x}_t + \tilde{y}_t, \eta_t, h_\mathcal{K}(\phi_t, \eta_t, \hat{x}_t + \tilde{y}_t)) \\
    \phi_{t+1} &= f_\mathcal{K}(\phi_t, \eta_t, \hat{x}_t + \tilde{y}_t) \\
    z_t &= [\hat{x}_t + \tilde{y}_t; h_\mathcal{K}(\phi_t, \eta_t, \hat{x}_t + \tilde{y}_t)].
\end{aligned}
\end{equation}
This system is exactly \eqref{eqn:cl-obsv-base-K} with the observer $\hat{x}_t = f(y_t, \eta_t, u_t)$ but for the addition of $\tilde{y}_t$ in the output $z_t$, which does not change the fact that it is contracting or Lipschitz. The result then follows by Theorem~\ref{thm:partial-converse}.

\subsubsection{Corollary~\ref{cor:partial-conv-certainty-equivalence}}
The closed-loop system $(\eta,d) \mapsto z$ consisting of \eqref{eqn:system} in feedback with $u 
= \mathcal{K}(\eta, y)$ \eqref{eqn:generic-ctrl} has dynamics
\begin{equation} \label{eqn:cl-perturbed}
\mathcal{K}:\; \left\{
\begin{aligned}
    x_{t+1} &= f(x_t, \eta_t, h_\mathcal{K}(\phi_t, \eta_t, h(x_t) + v_t)) + w_t \\
    \phi_{t+1} &= f_\mathcal{K}(\phi_t, \eta_t, h(x_t) + v_t) \\
    z_t &= [x_t; h_\mathcal{K}(\phi_t, \eta_t, h(x_t) + v_t)].
\end{aligned}\right.
\end{equation}
This system is contracting and Lipschitz by assumption. Take \eqref{eqn:cl-perturbed}, re-label $x$ as $\hat{x}$, and choose $v_t = \tilde{y}_t$, $w_t = L(\eta_t, \tilde{y}_t)$. Then the dynamics are exactly \eqref{eqn:cl-obsv-base-K} when the base controller and observer have the form \eqref{eqn:observer-linear-innovations}. The above transformation is Lipschitz (because $L$ is Lipschitz), hence if \eqref{eqn:cl-perturbed} is contracting and Lipschitz so too is \eqref{eqn:cl-obsv-base-K}. The result follows by Theorem~\ref{thm:partial-converse}.

%% file: main.bbl
\begin{thebibliography}{10}
\providecommand{\url}[1]{#1}
\csname url@samestyle\endcsname
\providecommand{\newblock}{\relax}
\providecommand{\bibinfo}[2]{#2}
\providecommand{\BIBentrySTDinterwordspacing}{\spaceskip=0pt\relax}
\providecommand{\BIBentryALTinterwordstretchfactor}{4}
\providecommand{\BIBentryALTinterwordspacing}{\spaceskip=\fontdimen2\font plus
\BIBentryALTinterwordstretchfactor\fontdimen3\font minus \fontdimen4\font\relax}
\providecommand{\BIBforeignlanguage}[2]{{%
\expandafter\ifx\csname l@#1\endcsname\relax
\typeout{** WARNING: IEEEtran.bst: No hyphenation pattern has been}%
\typeout{** loaded for the language `#1'. Using the pattern for}%
\typeout{** the default language instead.}%
\else
\language=\csname l@#1\endcsname
\fi
#2}}
\providecommand{\BIBdecl}{\relax}
\BIBdecl

\bibitem{Silver++2017}
D.~Silver \emph{et~al.}, ``Mastering the game of {G}o without human knowledge,'' \emph{Nature}, vol. 550, pp. 354--359, 2017.

\bibitem{tang2025deep}
C.~Tang, B.~Abbatematteo, J.~Hu, R.~Chandra, R.~Mart{\'\i}n-Mart{\'\i}n, and P.~Stone, ``Deep reinforcement learning for robotics: A survey of real-world successes,'' \emph{Annual Review of Control, Robotics, and Autonomous Systems}, vol.~8, no.~1, pp. 153--188, 2025.

\bibitem{Degrave++2022}
J.~Degrave \emph{et~al.}, ``Magnetic control of tokamak plasmas through deep reinforcement learning,'' \emph{Nature}, vol. 602, pp. 414--419, 2022.

\bibitem{Sutton+Barto2018}
R.~S. Sutton and A.~G. Barto, \emph{Reinforcement Learning: An Introduction}.\hskip 1em plus 0.5em minus 0.4em\relax The MIT Press, 2018.

\bibitem{williams1992simple}
R.~J. Williams, ``Simple statistical gradient-following algorithms for connectionist reinforcement learning,'' \emph{Machine learning}, vol.~8, no.~3, pp. 229--256, 1992.

\bibitem{meyn2022control}
S.~Meyn, \emph{Control systems and reinforcement learning}.\hskip 1em plus 0.5em minus 0.4em\relax Cambridge University Press, 2022.

\bibitem{Schulman++2017}
J.~Schulman, F.~Wolski, P.~Dhariwal, A.~Radford, and O.~Klimov, ``Proximal policy optimization algorithms,'' \emph{arXiv preprint arXiv:1707.06347}, 2017.

\bibitem{xu2022accelerated}
J.~Xu \emph{et~al.}, ``Accelerated policy learning with parallel differentiable simulation,'' in \emph{International Conference on Learning Representations (ICLR)}, 2022.

\bibitem{suh2022differentiable}
H.~J. Suh, M.~Simchowitz, K.~Zhang, and R.~Tedrake, ``Do differentiable simulators give better policy gradients?'' in \emph{International Conference on Machine Learning (ICML)}, 2022.

\bibitem{Freeman++2021}
C.~D. Freeman, E.~Frey, A.~Raichuk, S.~Girgin, I.~Mordatch, and O.~Bachem, ``Brax -- a differentiable physics engine for large scale rigid body simulation,'' in \emph{Advances in Neural Information Processing Systems (NeurIPS), Datasets and Benchmarks Track}, 2021.

\bibitem{howell2022dojo}
T.~A. Howell \emph{et~al.}, ``Dojo: A differentiable physics engine for robotics,'' \emph{arXiv preprint arXiv:2203.00806}, 2022.

\bibitem{Manchester++2026}
I.~R. Manchester, R.~Wang, and N.~H. Barbara, ``Neural networks in the loop: Learning with stability and robustness guarantees,'' \emph{Annual Review of Control, Robotics, and Autonomous Systems}, vol.~9, 2026.

\bibitem{Silver++2019}
T.~Silver, K.~Allen, J.~Tenenbaum, and L.~Kaelbling, ``Residual policy learning,'' \emph{arXiv preprint arXiv:1812.06298}, 2018.

\bibitem{Johannik++2019}
T.~Johannink \emph{et~al.}, ``Residual reinforcement learning for robot control,'' pp. 6023--6029, 2019.

\bibitem{Luo++2025}
J.~Y. Luo, Y.~Song, V.~Klemm, F.~Shi, D.~Scaramuzza, and M.~Hutter, ``Residual policy learning for perceptive quadruped control using differentiable simulation,'' in \emph{2025 IEEE International Conference on Robotics and Automation (ICRA)}.\hskip 1em plus 0.5em minus 0.4em\relax IEEE, 2025, pp. 1--8.

\bibitem{Youla++1976}
D.~C. Youla, J.~J. Bongiorno, and H.~A. Jabr, ``Modern {W}iener-{H}opf design of optimal controllers — {P}art {II}: The multivariable case,'' \emph{IEEE Transactions on Automatic Control}, vol.~21, pp. 319--338, 1976.

\bibitem{Anderson1998}
B.~D. Anderson, ``From {Y}oula-{K}ucera to identification, adaptive and nonlinear control,'' \emph{Automatica}, vol.~34, pp. 1485--1506, 12 1998.

\bibitem{Mahtout2020}
I.~Mahtout, F.~Navas, V.~Milanes, and F.~Nashashibi, ``Advances in {Y}oula-{K}u\v{c}era parametrization: A review,'' \emph{Annual Reviews in Control}, vol.~49, pp. 81--94, 2020.

\bibitem{Kucera1975}
V.~Ku\v{c}era, ``Stability of discrete linear feedback systems,'' \emph{IFAC Proceedings Volumes}, vol.~8, pp. 573--578, 8 1975.

\bibitem{Zhou++1996}
K.~Zhou, J.~C. Doyle, and K.~Glover, \emph{Robust and optimal control}.\hskip 1em plus 0.5em minus 0.4em\relax USA: Prentice-Hall, Inc., 1996.

\bibitem{Roberts++2011}
J.~W. Roberts, I.~R. Manchester, and R.~Tedrake, ``Feedback controller parameterizations for reinforcement learning,'' in \emph{IEEE Symposium on Adaptive Dynamic Programming and Reinforcement Learning (ADPRL)}, 2011, pp. 310--317.

\bibitem{Doyle1984}
J.~C. Doyle, ``Matrix interpolation theory and optimal control,'' 1984.

\bibitem{Zames1981}
G.~Zames, ``Feedback and optimal sensitivity: Model reference transformations, multiplicative seminorms, and approximate inverses,'' \emph{IEEE Transactions on automatic control}, vol.~26, no.~2, pp. 301--320, 1981.

\bibitem{Garcia+Morari1982}
C.~E. Garcia and M.~Morari, ``Internal model control. a unifying review and some new results,'' \emph{Industrial \& Engineering Chemistry Process Design and Development}, vol.~21, no.~2, pp. 308--323, 1982.

\bibitem{Goulart++2006}
P.~J. Goulart, E.~C. Kerrigan, and J.~M. MacIejowski, ``Optimization over state feedback policies for robust control with constraints,'' \emph{Automatica}, vol.~42, pp. 523--533, 4 2006.

\bibitem{simchowitz2020improper}
M.~Simchowitz, K.~Singh, and E.~Hazan, ``Improper learning for non-stochastic control,'' in \emph{COLT}, 2020, pp. 3320--3436.

\bibitem{Desoer+Liu1982}
C.~A. Desoer and R.-W. Liu, ``Global parametrization of feedback systems with nonlinear plants,'' \emph{System Control Letter}, vol.~1, pp. 249--251, 1982.

\bibitem{Desoer+Lin1983}
C.~A. Desoer and C.-A. Lin, ``Two-step compensation of nonlinear systems,'' \emph{Systems \& Control Letters}, vol.~3, no.~1, pp. 41--45, 1983.

\bibitem{sontag1989smooth}
E.~D. Sontag, ``Smooth stabilization implies coprime factorization,'' \emph{IEEE transactions on automatic control}, vol.~34, no.~4, pp. 435--443, 1989.

\bibitem{Tay+Moore1998}
T.~T. Tay and J.~B. Moore, ``Left coprime factorizations and a class of stabilizing controllers for nonlinear systems,'' in \emph{27th IEEE Conference on Decision and Control (CDC)}, 1988, pp. 449--454.

\bibitem{Paice+Moore1990}
A.~D. Paice and J.~B. Moore, ``On the {Y}oula-{K}ucera parametrization for nonlinear systems,'' \emph{Syst. Control Lett.}, vol.~14, pp. 121--129, 2 1990.

\bibitem{Paice+vanderSchaft1996}
A.~D. Paice and A.~J. V.~D. Schaft, ``The class of stabilizing nonlinear plant controller pairs,'' \emph{IEEE Transactions on Automatic Control}, vol.~41, pp. 634--645, 1996.

\bibitem{Fujimoto+Sugie2000}
K.~Fujimoto and T.~Sugie, ``Characterization of all nonlinear stabilizing controllers via observer-based kernel representations,'' \emph{Automatica}, vol.~36, pp. 1123--1135, 8 2000.

\bibitem{Lu1995}
W.~M. Lu, ``A state-space approach to parameterization of stabilizing controllers for nonlinear systems,'' \emph{IEEE Transactions on Automatic Control}, vol.~40, pp. 1576--1588, 1995.

\bibitem{Imura+Yoshikawa1997}
J.-i. Imura and T.~Yoshikawa, ``Parametrization of all stabilizing controllers of nonlinear systems,'' \emph{Systems \& Control Letters}, vol.~29, no.~4, pp. 207--213, 1997.

\bibitem{Lohmiller+Slotine1998}
W.~Lohmiller and J.~J.~E. Slotine, ``On contraction analysis for non-linear systems,'' \emph{Automatica}, vol.~34, pp. 683--696, 6 1998.

\bibitem{Wang+Manchester2023}
R.~Wang and I.~Manchester, ``Direct parameterization of {L}ipschitz-bounded deep networks,'' in \emph{International Conference on Machine Learning (ICML)}, 2023.

\bibitem{tobenkin2010convex}
M.~M. Tobenkin, I.~R. Manchester, J.~Wang, A.~Megretski, and R.~Tedrake, ``Convex optimization in identification of stable non-linear state space models,'' in \emph{49th IEEE Conference on Decision and Control (CDC)}, 2010, pp. 7232--7237.

\bibitem{tobenkin2017convex}
M.~M. Tobenkin, I.~R. Manchester, and A.~Megretski, ``Convex parameterizations and fidelity bounds for nonlinear identification and reduced-order modelling,'' \emph{IEEE Transactions on Automatic Control}, vol.~62, no.~7, pp. 3679--3686, 2017.

\bibitem{revay2020convex}
M.~Revay, R.~Wang, and I.~R. Manchester, ``A convex parameterization of robust recurrent neural networks,'' \emph{IEEE Control Systems Letters}, vol.~5, no.~4, pp. 1363--1368, 2020.

\bibitem{umenberger2018specialized}
J.~Umenberger and I.~R. Manchester, ``Specialized interior-point algorithm for stable nonlinear system identification,'' \emph{IEEE Transactions on Automatic Control}, vol.~64, no.~6, pp. 2442--2456, 2018.

\bibitem{Revay++2023}
M.~Revay, R.~Wang, and I.~R. Manchester, ``Recurrent equilibrium networks: Flexible dynamic models with guaranteed stability and robustness,'' \emph{IEEE Transactions on Automatic Control}, vol.~69, no.~5, pp. 2855--2870, 2024.

\bibitem{Wang+Manchester2022}
R.~Wang and I.~R. Manchester, ``{Youla-REN}: Learning nonlinear feedback policies with robust stability guarantees,'' in \emph{2022 American Control Conference (ACC)}.\hskip 1em plus 0.5em minus 0.4em\relax IEEE, 2022, pp. 2116--2123.

\bibitem{Wang++2022}
R.~Wang, N.~H. Barbara, M.~Revay, and I.~R. Manchester, ``Learning over all stabilizing nonlinear controllers for a partially-observed linear system,'' \emph{IEEE Control System Letter}, vol.~7, pp. 91--96, 2022.

\bibitem{Barbara++2023a}
N.~H. Barbara, R.~Wang, and I.~R. Manchester, ``Learning over contracting and {Lipschitz} closed-loops for partially-observed nonlinear systems,'' \emph{62nd IEEE Conference on Decision and Control}, pp. 1028--1033, 2023.

\bibitem{Kawano++2024}
Y.~Kawano, A.~J. van~der Schaft, and J.~M. Scherpen, ``{Y}oula--ku\v{c}era parameterization in contraction framework,'' \emph{IEEE Transactions on Automatic Control}, vol.~70, no.~3, pp. 1667--1682, 2025.

\bibitem{Furieri++2022}
L.~Furieri, C.~L. Galimberti, and G.~Ferrari-Trecate, ``Neural system level synthesis: Learning over all stabilizing policies for nonlinear systems,'' in \emph{IEEE Conference on Decision and Control}, 2022, pp. 2765--2770.

\bibitem{Furieri++2024}
------, ``Learning to boost the performance of stable nonlinear systems,'' \emph{IEEE Open Journal of Control Systems}, vol.~3, pp. 342--357, 2024.

\bibitem{Galimberti++2024}
C.~L. Galimberti, L.~Furieri, and G.~Ferrari-Trecate, ``Parametrizations of all stable closed-loop responses: From theory to neural network control design,'' \emph{Annual Reviews in Control}, vol.~60, p. 101012, 2025.

\bibitem{Megretski+Rantzer1997}
A.~Megretski and A.~Rantzer, ``System analysis via integral quadratic constraints,'' \emph{IEEE transactions on automatic control}, vol.~42, no.~6, pp. 819--830, 1997.

\bibitem{Zhou+Ren2001}
K.~Zhou and Z.~Ren, ``A new controller architecture for high performance, robust, and fault-tolerant control,'' \emph{IEEE Transactions on Automatic Control}, vol.~46, pp. 1613--1618, 2001.

\bibitem{Ho2020}
D.~Ho, ``A system level approach to discrete-time nonlinear systems,'' in \emph{IEEE American Control Conference (ACC)}, 2020, pp. 1625--1630.

\bibitem{Andrychowicz++2020}
M.~Andrychowicz \emph{et~al.}, ``What matters for on-policy deep actor-critic methods? a large-scale study,'' in \emph{International conference on learning representations (ICLR)}, 2021.

\bibitem{Barbara++2025a}
N.~H. Barbara, M.~Revay, R.~Wang, J.~Cheng, and I.~R. Manchester, ``Robustneuralnetworks.jl: a package for machine learning and data-driven control with certified robustness,'' \emph{Proc. JuliaCon Conf.}, vol.~7, no.~68, p. 163, 2025.

\bibitem{Wiedemann++2023}
N.~Wiedemann, V.~W{\"u}est, A.~Loquercio, M.~M{\"u}ller, D.~Floreano, and D.~Scaramuzza, ``Training efficient controllers via analytic policy gradient,'' in \emph{2023 IEEE International Conference on Robotics and Automation (ICRA)}.\hskip 1em plus 0.5em minus 0.4em\relax IEEE, 2023, pp. 1349--1356.

\bibitem{Barbara++2025b}
N.~H. Barbara, R.~Wang, and I.~R. Manchester, ``{R2DN}: Scalable parameterization of contracting and {L}ipschitz recurrent deep networks,'' \emph{arXiv preprint arXiv:2504.01250}, 2025.

\bibitem{andrieu2010uniting}
V.~Andrieu and C.~Prieur, ``Uniting two control {L}yapunov functions for affine systems,'' \emph{IEEE Transactions on Automatic Control}, vol.~55, no.~8, pp. 1923--1927, 2010.

\bibitem{Wang++2024}
R.~Wang, R.~Tóth, P.~J.~W. Koelewijn, and I.~R. Manchester, ``Virtual control contraction metrics: Convex nonlinear feedback design via behavioral embedding,'' \emph{International Journal of Robust Nonlinear Control}, vol.~34, pp. 7698--7721, 2024.

\bibitem{Rudin++2021}
N.~Rudin, D.~Hoeller, P.~Reist, and M.~Hutter, ``Learning to walk in minutes using massively parallel deep reinforcement learning,'' in \emph{Conference on Robot Learning (CoRL)}.\hskip 1em plus 0.5em minus 0.4em\relax PMLR, 2022, pp. 91--100.

\bibitem{Doyle1978}
J.~Doyle, ``Guaranteed margins for {LQG} regulators,'' \emph{IEEE Transactions on automatic Control}, vol.~23, no.~4, pp. 756--757, 1978.

\bibitem{Hochreiter+Schmidhuber1997}
S.~Hochreiter and J.~Schmidhuber, ``Long short-term memory,'' \emph{Neural Computation}, vol.~9, pp. 1735--1780, 11 1997.

\bibitem{Bakker2001}
B.~Bakker, ``Reinforcement learning with long short-term memory,'' in \emph{Advances in Neural Information Processing Systems (NeurIPS)}, 2001.

\bibitem{barbara2026thesis}
N.~H. Barbara, ``Parametrising neural feedback policies with stability and robustness guarantees,'' Ph.D. dissertation, University of Sydney, 2025.

\end{thebibliography}
